\documentclass[journal]{IEEEtran}

\usepackage{cite}

\ifCLASSINFOpdf
\usepackage[pdftex]{graphicx}
\graphicspath {{figures/}}
\else
\fi

\usepackage{amsmath}
\usepackage{algorithm}

\usepackage{array}

\ifCLASSOPTIONcompsoc
\usepackage[caption=false,font=normalsize,labelfont=sf,textfont=sf]{subfig}
\else
\usepackage[caption=false,font=footnotesize]{subfig}
\fi
\usepackage{url}
\usepackage{hyperref}

\usepackage[T1]{fontenc}
\usepackage[utf8x]{inputenc}

\usepackage{booktabs}
\usepackage{multirow}
\usepackage{cuted}
\usepackage{multicol}

\usepackage[dvipsnames]{xcolor}

\usepackage{tikz}
\usetikzlibrary{arrows, positioning, patterns}

\newcommand{\titlevar}{Lossless Coding of Light Fields\\based on 4D Minimum Rate Predictors}

\mathchardef\mhyphen="2D

\usepackage{anyfontsize}

\usepackage{amssymb}

\usepackage[noend]{algpseudocode}

\begin{document}
	\title{\titlevar}
	
	\author{João~M.~Santos,~\IEEEmembership{Student~Member,~IEEE,}
		Lucas~A.~Thomaz,~\IEEEmembership{Member,~IEEE,}
		Pedro~A.~A.~Assun\c{c}ão,~\IEEEmembership{Senior~Member,~IEEE,}
		Lu\'is~A.~da~Silva~Cruz,~\IEEEmembership{Senior Member,~IEEE},
		Lu\'is~T\'avora,
		and~S\'ergio~M.~M.~Faria,~\IEEEmembership{Senior~Member,~IEEE}
		\thanks{This work was supported by the Funda\c{c}\~ao para a Ci\^encia e a Tecnologia (FCT), Portugal, under PhD Grant SFRH/BD/114894/2016, Programa Operacional Regional do Centro, project PlenoISLA  POCI-01-0145-FEDER-028325 and  by FCT/MCTES through national funds and when applicable co-funded by EU funds under the project UIDB/EEA/50008/2020.}
		\thanks{The authors would like to thank Martin Briano for providing his implementation of CALIC.}%
		\thanks{João M. Santos and Luís A. da Silva Cruz are with Instituto de Telecomunica\c{c}\~oes and Department of Electrical and Computer Engineering,  Faculty  of Sciences and Technology,  University  of  Coimbra, Coimbra  3030-290, Portugal (e-mail: joao.santos@co.it.pt, luis.cruz@co.it.pt).}
		\thanks{Lucas A. Thomaz, Pedro A. A. Assun\c{c}\~ao and Sérgio M. M. Faria are with Instituto de Telecomunica\c{c}\~oes and  School of Technology and Management of Polytechnic of Leiria, Leiria 2411-901, Portugal (e-mail: lucas.thomaz@co.it.pt, amado@co.it.pt, sergio.faria@co.it.pt).}
		\thanks{Lu\'{i}s T\'avora is with School of Technology and Management of Polytechnic of Leiria, Leiria 2411-901, Portugal (e-mail: luis.tavora@ipleiria.pt).}
	}
	
	\markboth{IEEE Transactions on Image Processing}%
	{Santos \MakeLowercase{\textit{et al.}}: \titlevar}
	%

	
	
	\maketitle

	\begin{abstract}
		Common representations of light fields use four-dimensional data structures, where a given pixel is closely related not only to its spatial neighbours within the same view, but also to its angular neighbours, co-located in adjacent views. Such  structure presents increased redundancy between pixels, when compared with regular single-view images. Then, these redundancies are exploited to obtain compressed representations of the light field, using prediction algorithms specifically tailored to estimate pixel values based on both spatial and angular references. This paper proposes new encoding schemes which take advantage of the four-dimensional light field data structures to improve the coding performance of Minimum Rate Predictors. The proposed methods expand previous research on lossless coding beyond the current state-of-the-art. The experimental results, obtained using both traditional datasets and others more challenging, show bit-rate savings no smaller than 10\%, when compared with existing methods for lossless light field compression.
	\end{abstract}
	

	%
	\IEEEpeerreviewmaketitle

	\section{Introduction}
	
	\IEEEPARstart{L}{ight} field (LF) imaging, initially proposed by Lippmann in 1908~\cite{Lippmann_1908_JPTA}, promises to provide superior immersive multimedia experiences for users and content producers. Light fields fully characterise the flow of light in every direction and point in space. With appropriate sampling, this function can be translated to imaging data that can represent both traditional 2D spatial information and also the directionality of light rays, \textit{i.e.}, angular information, reaching the camera sensors. Applications of this technology range from the entertainment industry, such as virtual and augmented reality~\cite{Wang_2015_JoDT_AugmentedReality3DDisplaysWithMicroIntegralImaging} and 3D movies and television~\cite{Arai_2013_OE_IntegralThree-dimensionalTelevisionWithVideoSystemUsingPixel-offsetMethod}, to more critical tasks as medical imaging~\cite{Shin_2010_OL_Three-dimensionalOpticalMicroscopyUsingAxiallyDistributedImageSensing, Pereira_2020_BSPaC_DermoscopicSkinLesionImageSegmentationBasedOnLocalBinaryPatternClustering, Pereira_2020_BSPaC_SkinLesionClassificationEnhancementUsingBorderLineFatures}, for instance. These various applications are only possible due to the characteristics of LFs, which represent increased visual information when compared to traditional 2D images. Such extra information enables a panoply of post-processing operations, like the extraction of depth maps~\cite{Zhou_2020_ITIP_UnsupervisedMonocularDepthEstimationFromLightFieldImage, Mishiba_2020_ITIP_FastDepthEstimationforLightFieldCameras, Lourenco_2018_IICoIP_SilhouetteEnhancementInLightFieldDisparityEstimationUsingTheStructureTensor}, or image rendering with different viewing perspectives~\cite{Jung_2020_IToM_FlexiblyConnectableLightFieldSystemForFreeViewExploration}, or post-capture refocusing~\cite{Wang_2018_ISPL_SelectiveLightFieldRefocusingForCameraArraysUsingBokehRenderingAndSuperresolution}.
	
	Developments in LF acquisition systems technology are expected to make LF imaging available to a broader range of users, both industrial and domestic. However, one of the potential deterrents of adoption of LF imaging is the huge amount of storage capacity and bandwidth required to handle all the data required by this technology. In consideration of this, both the Joint Photographic Experts Group (JPEG)~\cite{WG1N74014_2017_JPEGPlenoCallforProposalsonLightFieldCoding} and the Moving Picture Experts Group (MPEG)~\cite{WG11N16352_2017} started standardisation initiatives with the objective of developing efficient encoding techniques of LFs, ranging from lossy to lossless compression.
	
	Light field compression has been the subject of intensive research by the scientific community, but most of this effort has been directed towards lossy coding, as evidenced by the extensive survey of~\cite{Conti_2020_IA_DenseLightFieldCodingaSurvey}. However, not all applications can afford the loss of information, in particular, applications such as medical imaging or precision measurements in industry need high fidelity in the reconstruction of images, requiring lossless and near-lossless performance. Lossless coding, on the other hand, is a less explored topic in LF compression. Despite the JPEG-Pleno call for proposals~\cite{WG1N74014_2017_JPEGPlenoCallforProposalsonLightFieldCoding} asking for both lossy and lossless coding solutions, from the selected algorithms for the standard only WaSP~\cite{Astola_2018_EWVIPE_WaSPHierarchicalWarpingMergingandSparsePredictionforLightFieldImageCompression} is currently capable of achieving lossless compression. Even so, in the JPEG-Pleno standardisation activity experiment~\cite{WG1N82044_2019_VM2Lossless} it was shown that a non-standard encoder~\cite{Santos_2019_DCC}, based on Minimum Rate Predictors (MRP)~\cite{Matsuda_2007_SaCiJ_ALosslessCodingSchemeUsingAdaptivePredictorsandArithmeticCodeOptimizedforEachImage} achieves higher lossless compression efficiency.
	The underlying principles of MRP consist of estimating  the amount of information conveyed by prediction errors resulting from predictors, which are specifically designed to minimise the rate rather than the prediction error itself, as usually done in lossy encoding schemes. 
	
	LFs are usually represented as 4D functions, with two spatial dimensions accounting for the 2D spatial information, also present in regular images $ (v, u) $, and two angular dimensions accounting for the different viewpoints that are also captured in LFs $ (t, s) $. 4D LFs present increasing redundancy, when compared with 2D images, that can be exploited to provide higher compression efficiency than that provided by encoders which cannot exploit all this extra information. This paper exploits these redundancies present in the 4D LF with the proposal of efficient methods of 4D prediction and partition for the MRP codec. Two new modes and the prediction mode proposed by the authors in~\cite{Santos_2019_DCC} are studied in this paper, each with its own characteristics in terms of the trade-off between complexity and compression efficiency. Although M-MRP was already introduced in \cite{Santos_2019_DCC}, in the current paper its performance undergoes a more thorough evaluation, with updated test conditions, following the common test conditions of JPEG-Pleno (JPEG-Pleno CTC) standardisation initiative~\cite{JPEG-CTC}, and using a wider variety of datasets. M-MRP is presented alongside the new prediction modes so that their commonalities and differences are highlighted. The three modes studied in this paper are entitled the 4D-MRP, Dual-Tree 4D-MRP, and multiple-reference MRP (M-MRP) \cite{Santos_2019_DCC}. In the first, an hexadecatree partitioning structure is used to enable four-dimensional prediction and 4D blocks as coding units. In the second, the angular and spatial partitioning are separate, such that each 4D block can be partitioned into four blocks in either the angular dimensions $ (t, s) $ or spatial dimensions $ (v, u) $,  rather than being  partitioned  in  16  blocks  as  in  4D-MRP. Finally, in M-MRP only 2D partitions are used, using the same 4D prediction of the other modes but losing the benefits of 4D partition. The main contributions of this paper can be summarised as follows:
	\begin{itemize}
		\item Development of a unified native 4D framework for lossless compression of light fields based on Minimum Rate Predictors.
		\item Design of two novel prediction modes for the native 4D lossless compression framework, namely 4D-MRP and DT-4D-MRP.
		\item Extensive evaluation and comparison of the three prediction modes in the context of the JPEG-Pleno CTC, amongst each other and with state-of-the-art encoders.
		\item Assessment of the three prediction modes in a medical context, using the Skin Lesion Light-fields (SKINL2) dataset~\cite{Faria_2019_24AICotIEiMaBSE_LightFieldImageDatasetofSkinLesions}.
	\end{itemize}
	
	The experimental assessment clearly ascertains the superior performance of the proposed methods in comparison with other state-of-the-art encoders, such as HEVC~\cite{Sullivan_2012_HEVC}, VVC~\cite{Bross_2021_PI_VVC}, and JPEG XL~\cite{Alakuijala_2019_JPEGXL}. In fact, the presented methods surpass other encoders by up to 32\%, in terms of bitrate savings.
	
	The remainder of the paper is organised as follows, Section~\ref{sec:4d-lf} presents the theory behind the 4D representation of LFs, Section~\ref{sec:related_work} gives an overview of other lossless compression of LFs works, Section~\ref{sec:4d-mrp} describes the proposed encoders, in Section~\ref{sec:experimental-results} the experimental results are presented and analysed, and, finally, Section~\ref{sec:conclusion} discusses the outcomes of the paper.
	
	\section{Background}
	
	\subsection{4D Light Field Representation} \label{sec:4d-lf}
	
	Light fields are sampled representations of the plenoptic function, which characterises the intensity $\mathcal{L}$ of light rays at every point in space $ (x, y, z) $, travelling in the $ (\theta, \phi) $ direction, with the wavelength $ \lambda $, at time instant $ t $ \cite{Zhang_2006_LFS_LightFieldSampling}. This is a 7D function represented by
	\begin{equation}
		\mathcal{L} = P\left(x, y, x, \theta, \phi, \lambda, t \right).
	\end{equation}
	
\begin{figure}[t]
		\centering
		\resizebox{\columnwidth}{!}{\input{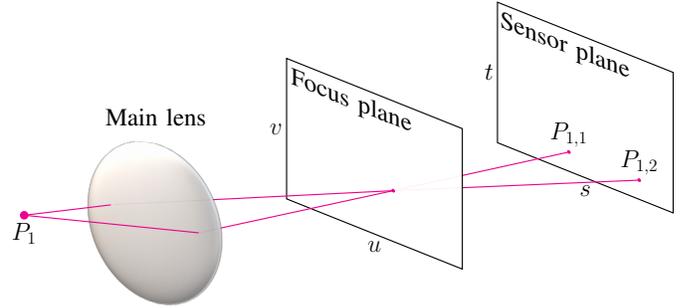}}
		\caption{Simplified example of the 4D LF parametrisation in a plenoptic camera.}
		\label{fig:4d-lf-parametrisation}
	\end{figure}

	However, in practical systems full representation of the plenoptic function is not necessary, thus a simplified version is normally used. This simplified version also requires more reasonable computational effort to sample due to inherent data reduction. Several different representations have been proposed in the literature each with distinct characteristics~\cite{Levoy_1996_CoCGaIT_LightFieldRendering}. This paper employs the common 4D representation of LFs, which reduces the seven dimensions of the plenoptic function to four, based on the following three assumptions:
	\begin{enumerate}
		\item the wavelength $ \lambda $ can be simplified to the three RGB channels normally captured by cameras (that represent specific ranges of wavelength).
		\item The transmission medium does not introduce attenuation, so the radiance along a light ray path remains constant. Therefore the scene information can be captured by projection onto any selected surface and the $ z $ dimension can be eliminated.
		\item The scene is static during the acquisition interval thus, the time dimension $ t $ can be disregarded.
	\end{enumerate}
	Under these assumptions the 7D plenoptic function can be simplified to a 4D function:  
	\begin{equation}\label{eq:4d-lf}
		\mathcal{L}_{4D} = L(t, s, v, u).
	\end{equation}
	A 4D parametrisation of a light field using two planes is shown in Figure~\ref{fig:4d-lf-parametrisation}, where a light ray direction is defined by the coordinates of its intersection points with the $ (v, u) $ and $ (t, s) $ planes. Light rays coming from the same scene point with different directions pass through the same focal point in the $ (v, u) $ plane and then are uniquely identified by the coordinates in the $ (t, s) $ plane. In other words, for each point $ (v, u) $ the camera captures the intensity of light rays with different directions $ (t, s) $. In this context, $ (t, s) $ represent the coordinates of sub-aperture images (SAI) in the 4D LF, and $ (v, u) $ represent pixel coordinates in a SAI. The two planes represented in Figure~\ref{fig:4d-lf-parametrisation} are discretised allowing for only a finite number of viewpoints and light rays to be recorded.
	
	Several processes for capturing LFs exist, as described by Zhou in~\cite{Zhou_2011_IToIP_ComputationalCamerasConvergenceofOpticsandProcessing}. For the purposes of this paper, single sensor LF cameras and array of cameras are considered. Considering the former, two main types of LF cameras can be found: the focused camera, also known as plenoptic 2.0 and the  unfocused camera, also known as plenoptic 1.0. Several publicly available LF datasets can be found in~\cite{JPEG-CTC} and~\cite{Faria_2019_24AICotIEiMaBSE_LightFieldImageDatasetofSkinLesions}. Despite the differences in the acquisition processes, LFs can always be represented as a 4D data structure, comprising a set of SAIs with the same spatial resolution, each one representing a different viewpoint of the same visual scene. Usually these viewpoints are captured along the vertical and horizontal directions giving rise to disparity in both directions. The described 4D LF format is also adopted in JPEG-Pleno encoding framework and in their standard encoders~\cite{Astola_2018_EWVIPE_WaSPHierarchicalWarpingMergingandSparsePredictionforLightFieldImageCompression, Carvalho_2018_IICoIPI_A4DDCTBasedLensletLightFieldCodec}, which use the 4D LFs in the SAI domain~\cite{JPEG-CTC}.
	
	\subsection{Coding of Light Fields} \label{sec:related_work}
	
	Lossless compression techniques can be grouped according to LF representation format: lenslet sensor data, SAI or epipolar images (EPI). Considering the first category, encoders compress the raw `RGGB' Bayer image obtained from the camera sensor, before the pre-processing steps, as in~\cite{Dansereau_2013_CVaPR_DecodingCalibrationAndRectificationForLenseletBasedPlenopticCameras}. In this case, the metadata of the camera must be taken into account when calculating compression efficiency, as this information is required to extract images from the LF. In~\cite{Perra_2015_IICASaSPI_LosslessPlenopticImageCompressionUsingAdaptiveBlockDifferentialPrediction} the author proposes to split the image into four colour planes that are entropy analysed to discover the horizontal and vertical displacement between the micro images. Later, a sequential differential prediction process is performed, and the residue is encoded using a Lempel-Ziv-Markov chain algorithm~\cite{Salomon_2007_DCC}. The authors in~\cite{Tabus_2017_ESPCE_MicrolensImageSparseModellingforLosslessCompressionofPlenopticCameraSensorImages} propose a different approach, where the nine closest causal micro-images are used to predict the current micro-image. This is achieved by designing a minimum description length optimal sparse predictor for each micro-image, where the non-zero coefficients of the predictor are selected by using a binary mask, which also needs to be transmitted. More recently, in \cite{Tabus_2021_IAccess} the authors propose a codec for plenoptic camera sensor images dubbed sparse relevant regressors and contexts. The codec splits the lenslet image in patches, each corresponding to a micro-lens. The encoder exploit the intra- and inter-patch correlation, induced by the cameras bayer pattern and the micro-lenses array, by designing a sparse predictor for each of the patches pixels. Miyazawa \textit{et al.} use a modified version of MRP to encode the `RGGB'  raw image from the camera sensor~\cite{Miyazawa_2018_IWoAITI_LosslessCodingofLightFieldCameraDataCapturedwithaMicroLensArrayandaColorFilter}. In their work, two types of predictors are designed, Type-I uses reference samples from the causal area of the target sample (which might encompass regions from other micro-images), this predictor allows the exploitation of image correlations expressed in both the spatial and spectral domains. The Type-II predictor exploits inter micro-images redundancies, thus the reference samples are placed in previously encoded micro-images centred around the same pixel location in all reference micro-images. Several predictors are designed and selected on a hexagonal block partitioning basis. In \cite{Schiopu_2018_IICoIPI_MacroPixelPredictionBasedonConvolutionalNeuralNetworksforLosslessCompressionofLightFieldImages}, Schiopu proposed to encode macro-pixels using deep learning based prediction, which was later extended in \cite{Schiopu_2019_AToSaIP_DeepLearningBasedMacroPixelSynthesisandLosslessCodingofLightFieldImages}. These works employ deep learning to both compress and synthesise the LF images. A set of reference views are selected and losslessly encoded using a REP-CNN~\cite{Schiopu_2018_EL_ResidualErrorPredictionBasedonDeepLearningForLosslessImageCompression}, then the whole LF is synthesised based on the reference views. The synthesised LF is used as a base for a deep-learning based method that encodes the details of the LF image.
	
	Despite the previous methods being specifically tailored for lenslet formats, the major focus of LF lossless coding has been on compressing SAIs. In~\cite{Helin_2016_3TTVTaDo3V3_SparseModellingandPredictiveCodingofSubapertureImagesforLosslessPlenopticImageCompression}, \cite{Helin_2017_IJoSTiSP_MinimumDescriptionLengthSparseModelingandRegionMergingforLosslessPlenopticImageCompression}, and \cite{Tabus_2017_IICoIPI_LossyCompressionofLensletImagesfromPlenopticCamerasCombiningSparsePredictiveCodingandJPEG2000} Helin \textit{et al.} partition the views into regions, which are expanded from the central view using vertical and horizontal displacements that need to be transmitted. The central view is losslessly encoded with JPEG2000, the side views use the previously encoded ones as reference with a region specific sparse linear predictor. In~\cite{Schiopu_2017_ICoIPTTaAI_SubapertureImageSegmentationforLosslessCompression} and \cite{Schiopu_2017_3DTVConf_LosslessCompressionofSubapertureImagesUsingContextModeling} Schiopu \textit{et al.} propose to use segmentation, after the predictive coding of the LF views, to improve the context modelling of the residuals for the entropy coder. Rizkallah \textit{et al.} in~\cite{Rizkallah_2019_DCCD_GraphBasedSpatioAngularPredictionforQuasiLosslessCompressionofLightFields} and \cite{Rizkallah_2020_IToIP_PredictionandSamplingwithLocalGraphTransformsforQuasiLosslessLightFieldCompression} propose to use graph-based transforms for quasi-lossless compression of LFs.
	
	Finally, few works are dedicated to the lossless compression of LFs in the EPI format. Mukati and Forchhammer propose to adapt the CALIC codec~\cite{Wu_1997_IToC_ContextBasedAdaptiveLosslessImageCoding} to compress EPI images~\cite{Mukati_2020_2DCCD_EPICContextAdaptiveLosslessLightFieldCompressionUsingEpipolarPlaneImages}. The CALIC GAP predictor is improved so that it can detect the EPI slope, thus the name `EPI slope based predictor' (ESP), in order to appropriately predict the pixel intensities along each epipolar line. This proposal is shown to increase the compression performance of CALIC for LFs by 0.8 bits-per-pixel (bpp). This work is further expanded in~\cite{Mukati_2020_IAccess}, which increases the performance of the encoder by 5.3\%.
	
	The authors' previous works~\cite{Santos_2018_JoVCaIR_LosslessCodingofLightFieldImagesBasedonMinimumRatePredictors} and \cite{Santos_2017_ICoIPTTaAI_LosslessLightFieldCompressionUsingReversibleColourTransformations} fall in all previous three categories, by comparing the performance of various state-of-the-art encoders on various LF data arrangements, with focus on Minimum Rate Predictors encoder (MRP)~\cite{Santos_2016_PCSP_CompressionofMedicalImagesUsingMRPwithBiDirectionalPredictionandHistogramPacking}. In these works various reversible colour transforms are tested to increase the lossless compression performance of such encoders.
	
	Another work worth mentioning is~\cite{Thomaz_2019_EELFIWE_VisuallyLosslessCompressionofLightFields}, resulting from a JPEG-Pleno experiment intended to compare the lossless and near-lossless performance of upcoming JPEG-Pleno standard codecs, the WaSP and MuLE (which is currently lossy only) encoders to M-MRP~\cite{Santos_2019_DCC}. MuLE and WaSP present lower bit-rate to obtain near-lossless compression, when compared with the lossless bit-rate of the M-MRP. However, when targeting lossless compression, the JPEG-Pleno standard codecs are unable to keep up with the M-MRP performance, in some cases achieving twice the bit-rate of M-MRP.

	\section{4D Prediction Modes for MRP} \label{sec:4d-mrp}
	
	\subsection{The MRP coding algorithm} \label{ssec:mrp-algorithm}
	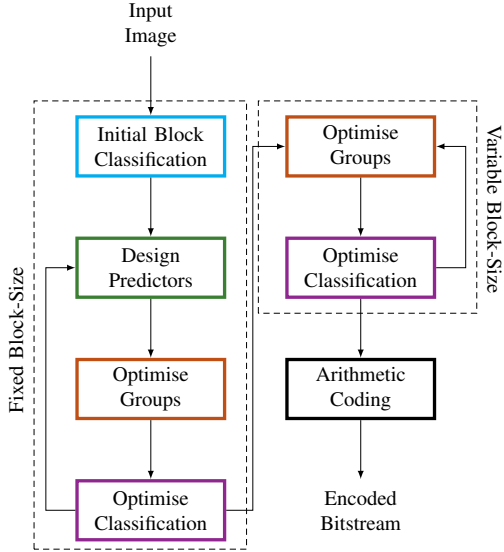
\begin{figure}[t]
		\centering
		\resizebox{0.75\columnwidth}{!}{\begin{tikzpicture}[anchor=south west, inner sep=0pt]
	\tikzset{block/.style= {draw, rectangle, ultra thick, align=center,minimum width=2.5cm,minimum height=1cm},};

	\node [block, draw=cyan] (init) {Initial Block\\Classification};
	\node [block, draw=none, above= of init, minimum width=0cm] (begin) {Input\\Image};
	\node [block, draw=OliveGreen, below=of init] (design) {Design\\Predictors};
	\node [block, draw=Bittersweet, below=of design] (opt-groups-1) {Optimise\\Groups};
	\node [block, draw=Plum, below=of opt-groups-1] (opt-classification-1) {Optimise\\Classification};
	\node [block, draw=Bittersweet, right=of init] (opt-groups-2) {Optimise\\Groups};
	\node [block, draw=Plum, below=of opt-groups-2] (opt-classification-2) {Optimise\\Classification};
	\node [block, below=of opt-classification-2] (ac) {Arithmetic\\Coding};
	\node [block, draw=none, below= of ac, minimum width=0cm] (end) {Encoded\\Bitstream};

	\draw [-latex] (begin) -- (init);
	\draw [-latex] (init) -- (design);
	\draw [-latex] (design) -- (opt-groups-1);
	\draw [-latex] (opt-groups-1) -- (opt-classification-1);
	\draw [-latex] (opt-classification-1.west) -- ++(-0.5, 0) node{} |- (design.west);
	\draw [-latex] (opt-classification-1.east) -- ++(0.45, 0) node{} |- (opt-groups-2.west);
	\draw [-latex] (opt-groups-2) -- (opt-classification-2);
	\draw [-latex] (opt-classification-2.east) -- ++(0.5,0) node{} |- (opt-groups-2.east);
	\draw [-latex] (opt-classification-2) -- (ac);
	\draw [-latex] (ac) -- (end);

	\draw [xshift=1.5 cm, yshift=-6.1 cm, densely dashed] (-2.2, -0.2) -- node[left, rotate=90, yshift=0.3cm, xshift=1.cm] {Fixed Block-Size} ++ (0, 7.6cm) -- ++(3.6, 0) -- ++(0, -7.6cm) -- cycle;
	\draw [xshift= 1.6cm, yshift=-4.1 cm, densely dashed] (1.5, 1.8) -- ++ (0, 3.6) -- ++ (3.7,0) -- node[right, rotate=-90, yshift=0.3cm, xshift=-1.4cm] {Variable Block-Size} ++ (0, -3.6) -- cycle;
\end{tikzpicture}}
		\caption{Functional diagram of the MRP family algorithms~\cite{Santos_2017_ICoIPTTaAI_LosslessLightFieldCompressionUsingReversibleColourTransformations}.}
		\label{fig:mrp-fluxogram}
	\end{figure}
	
	To cope with the content diversity of generic images, MRP uses an adaptive prediction scheme for each image. As for the prediction residue, the algorithm employs entropy coding based on context modelling. Such prediction residues are sorted into one of several predetermined groups, \textit{i.e.} contexts, depending on their characteristics. The set of residues pertaining to a given group are later represented by a single generalised Gaussian probability density function~\cite{Matsuda_2007_SaCiJ_ALosslessCodingSchemeUsingAdaptivePredictorsandArithmeticCodeOptimizedforEachImage} whose parameters differ from those of the other groups.
	
	Mathematically, MRP is formulated as the search for linear predictors, \textit{i.e.} classes, that minimise the number of bits used to encode the prediction errors and associated information. Classes are characterised by linear prediction models each with a set of coefficients. As described in~\cite{Matsuda_2007_SaCiJ_ALosslessCodingSchemeUsingAdaptivePredictorsandArithmeticCodeOptimizedforEachImage}, the information associated with the prediction errors $ e $ in a given image region $ R $, is given by
	\begin{equation}\label{eq:mrp-information}
		I(R) = \sum_{n = 0}^{N - 1}\left\lbrace - \sum_{p_0 \in g_n} \log_2 \alpha_n + \frac{\log_2 \epsilon}{2} \cdot \sum_{p_0 \in g_n} \frac{e^2}{\sigma_n^2} \right\rbrace,
	\end{equation}
	where $ I(R) $ indicates the estimate of the total amount of information measured in bits, $ N $ is the number of groups associated with the context modelling, $ g_n $ are the pixels in the $ n^{th} $ group, $ \sigma_n $ is the variance of the prediction error $ e $ of the $ n^{th} $ group, $ \epsilon $ is Euler's number, and $ \alpha_n =\frac{\Delta e }{\sqrt{2 \pi \sigma_n^2}} $, with $ \Delta e $ representing a sufficiently small quantisation step-size of $ e $. The objective of the encoding procedure is to minimise $ I(R) $, by properly determining the sets of linear prediction coefficients for the $ M $ classes, that are used to compute the prediction and associated residuals. In practical terms, the regions in Equation~\ref{eq:mrp-information} are the partition blocks and the actual minimisation target is given by a cost function accounting for the prediction residuals and associated side information. The classes linear prediction models in MRP are adapted to each image characteristics, through the use of a variable block size partitioning scheme in the iterative minimisation of the cost function. In general, the MRP algorithm processing flow can be represented by the block diagram shown in Figure~\ref{fig:mrp-fluxogram}, which is further described by Algorithm~\ref{alg:mrp}, where each colour represents the same operation in both the figure and the algorithm. The modules include the three main stages, \textit{Fixed Block-Size}, \textit{Variable Block-Size}, and \textit{Arithmetic Coding}.
	\begin{algorithm}[ht!]
		\caption{Minimum rate predictors high level algorithm.}
		\label{alg:mrp}
		\begin{algorithmic}[1]
			\State \textbf{Input:} image
			
			\State Apply pre-processing
			
			\State \textcolor{cyan}{\textbf{\#Initial block classification}}
			\State Sort and classify of $ 8 \times 8 $ blocks by pixel variance
			
			\State \textbf{\#First optimisation loop using fixed block size}
			\For {1 \textbf{to} MAX\_ITERATIONS}
			
			\State \textcolor{OliveGreen}{\textbf{\#Design predictors}}
			\For {each class}
			\State Compute the prediction coefficients ($ a_{m_i} $)
			\EndFor
			\State Calculate prediction residue encoding cost ($ B_r $)
			
			\State \textcolor{Bittersweet}{\textbf{\#Optimise Groups}}
			\For {each class}
			\State Compute $ C $ quant. thresholds to minimise $ B_r $
			\EndFor
			\State Calculate $ B_r $
			
			\State \textcolor{Plum}{\textbf{\#Optimise classification}}
			\For {each $ 8 \times 8 $ block}
			\State Move neighbouring blocks classes to front of table
			\State Select class that minimises $ B_r $
			\EndFor
			\color{black}
			\State Calculate $ B_r $
			\If{10 iterations without improvement} \State \textbf{end} for loop \EndIf
			\EndFor
			
			\State \textbf{\#Second optimisation loop using variable block size}
			\For {1 \textbf{to} MAX\_ITERATIONS}
			\State Calculate prediction coefficients encoding cost ($ B_a $)
			\State \textcolor{Bittersweet}{\textbf{\#Optimise Groups}}
			\For {each class}
			\State Compute $ C $ quant. thresholds to minimise $ B_r $
			\State Update shape parameter in probability models
			\EndFor
			\State Calculate quantisation thresholds cost ($ B_t $)
			\State\textcolor{Plum}{\textbf{\#Optimise classification}}
			\For {Each $ 32 \times 32 $ block}
			\State \Call{OptimiseClass}{}
			\EndFor
			\State Calculate cost $ B_m $
			\State Calculate cost $ J $
			
			\If{10 iterations without improvement} \State \textbf{end} for loop \EndIf
			\EndFor
			
			\State Remove non-utilised classes
			\State Run arithmetic coding
			\State \textbf{Output:} Encoded bitstream
			\State 
			\Procedure{OptimiseClass}{}
			\State Move neighbouring blocks classes to front of table
			\State Select class that minimises $ B_r $
			
			\If{$ \mathrm{level} > 0 $}
			\State Calculate cost of not partitioning block ($ J_{1} $)
			
			\State Partition block in quadtree fashion
			\For{each resulting block}
			\State \Call{Optimise Class}{}
			\EndFor
			\State Calculate sum of cost of partitioned blocks ($ J_{2} $)
			
			\If{$J_2 < J_1 $} \State Partition block \EndIf
			\EndIf
			\State \textbf{Return:} Cost, partition structure and class selection
			\EndProcedure
		\end{algorithmic}
	\end{algorithm}
	
	Prior to the actual encoding, a pre-processing step, consisting in applying a Reversible Colour Transform (RCT) and an histogram packing step, is applied to the LF. The RCT, which was first combined with MRP in \cite{Santos_2017_ICoIPTTaAI_LosslessLightFieldCompressionUsingReversibleColourTransformations}, aims to lower the RGB inter-component correlation to improve the coding performance, the transform is defined in Equation~\ref{eq:rct}, where $ \lfloor . \rfloor $ represents the floor operation. The histogram packing, first introduced to MRP in~\cite{Santos_2016_PCSP_CompressionofMedicalImagesUsingMRPwithBiDirectionalPredictionandHistogramPacking}, is used to reduce the sparseness of the LF histogram, improving the performance of lossless prediction based encoders, provided that it produces a LF image with lower total variation than the original LF.
	\begin{equation}\label{eq:rct}
		\left.
		\begin{aligned}
			& Y = \left\lfloor{\frac{R + 2G + B}{4}}\right\rfloor \\
			& C_u = B - G \\
			& C_v = R - G
		\end{aligned}
		\right\}
		\ \Leftrightarrow \ 
		\left\{
		\begin{aligned}
			& R = C_v + G \\
			& G = Y - \left\lfloor{\frac{C_u + C_v}{4}}\right\rfloor \\
			& B = C_u - G
		\end{aligned}
		\right.
	\end{equation}
	In the \textit{Fixed Block-Size} stage blocks of $ 8 \times 8 $ pixels are used to calculate the coefficients of the $ m $ linear prediction models that define the corresponding classes. First the blocks are sorted according to their pixel variance and distributed by the classes in an increasing variance order to initialize the algorithm. For each class $ m $, the coefficients ($ a_{m_i} $) of the corresponding linear prediction model are determined in module \textit{Design Classes}, by solving a set of Yule-Walker equations, which uses the pixels of all blocks associated with that class. Then, in \textit{Optimise Groups} the thresholds for the quantisation of the prediction residuals context, represented by $ C $, are computed. The quantisation of the value $ C $ generates 16 different contexts, given by 15 thresholds that are calculated on a class basis. This means that all the blocks of a given class are used to compute these thresholds. Due to the dynamic nature of the thresholds, these are explicitly transmitted in the bitstream. Finally, in \textit{Optimise Classification} the $ 8 \times 8 $ blocks are reclassified into the newly computed classes linear prediction models, with the objective of minimising the optimisation target. Each class is tested in order to find the one that minimises the block prediction residuals cost. Instead of directly transmitting the actual class index for each block, the MRP uses a lookup table that is updated as each block is processed by using a move-to-front method. The class index associated with the upper, left, and upper right neighbour blocks are placed at the top of the lookup table, in this order. In this manner, as neighbour blocks are correlated, the indices to transmit should generally have small values, which leads to a more efficient compression of these indices (\textit{i.e.}, different classes can be represented by the same lookup table index). Additionally, due to the optimisation process, classes of neighbouring blocks are favoured, especially in flat areas. In this first optimisation loop, the minimisation target is the prediction residuals encoding cost estimated by:
	\begin{equation}\label{eq:mrp-br}
		B_r = \sum\limits_{p_0}L\left( e|\hat{s}(0), n \right),
	\end{equation}
	where $ L(e|\hat{s}(0), n) $ is the encoding cost of the prediction error, which depends on the probability density estimate of the $ n \mhyphen th $ group, assumed to conform with a Gaussian distribution. By minimizing the cost $ B_r $, that is, the encoding costs of the prediction errors, the method implicitly optimizes the information $ I(R) $ in a given region (or blocks), therefore minimizing Equation~\ref{eq:mrp-information}.
	
	The \textit{Variable Block-Size} stage refines the operations of the previous optimisation loop, by taking into account the cost of all signalling flags, using variable block size and fitting the probability models. In addition to the operations of the previous loop, \textit{Optimise Groups} updates the shape parameter of generalised Gaussian probability density functions to get a good fit for the prediction error of each group. In \textit{Optimise Classification}, the classes linear prediction models designed in the previous stage are used to optimise the block classification resorting to variable block sizes, partitioned in a quadtree fashion. The block sizes range from $ 32 \times 32 $ pixels down to $ 2 \times 2 $ pixels, for quadtree levels from $ 4 $ down to $ 0 $, respectively. The optimal partition is selected so that at each level, \textit{i.e.} block size, the minimum between the current block cost ($ J_1 $) and the sum of the costs of the sub-blocks ( $ J_2 $) is selected. The target cost function of the second loop is defined as a sum of several costs expressed in bits:
	\begin{equation}\label{eq:mrp-cost}
		J = B_a + B_m + B_t + B_r,
	\end{equation}
	where $ B_a $, $ B_m $, and $ B_t $ are the encoding costs of the prediction coefficients, class selection, and context modelling threshold values, respectively, and $ B_r $ is defined in Equation~\ref{eq:mrp-br}.
	
	The resulting information from these two optimisation loops is then encoded using an arithmetic encoder~\cite{Matsuda_2007_SaCiJ_DesignandEvaluationofMinimumRatePredictorsforLosslessImageCoding}. More details on the LF encoding based on the MRP algorithm can be found in~\cite{Santos_2017_ICoIPTTaAI_LosslessLightFieldCompressionUsingReversibleColourTransformations, Santos_2018_JoVCaIR_LosslessCodingofLightFieldImagesBasedonMinimumRatePredictors}.
	
\begin{figure}[t]
		\centering
		\resizebox{0.9\columnwidth}{!}{\input{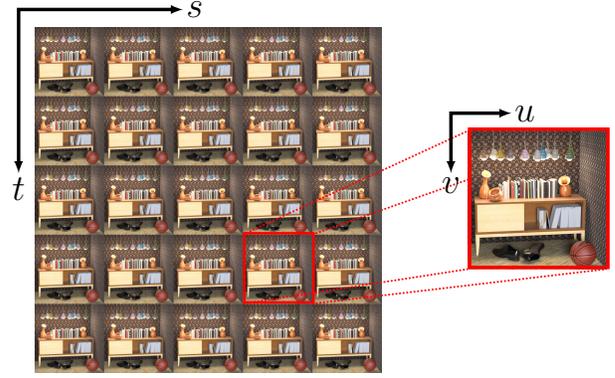}}
		\caption{Example of a 4D light field in a sub-aperture image array arrangement, Sideboard image from~\cite{Honauer_2016_ACoCV_ADatasetandEvaluationMethodologyforDepthEstimationon4dLightFields}.}
		\label{fig:4d-lf}
	\end{figure}

\begin{figure}[t]
		\centering
		\includegraphics[width=0.8\linewidth, trim={0 256 256 0}, clip]{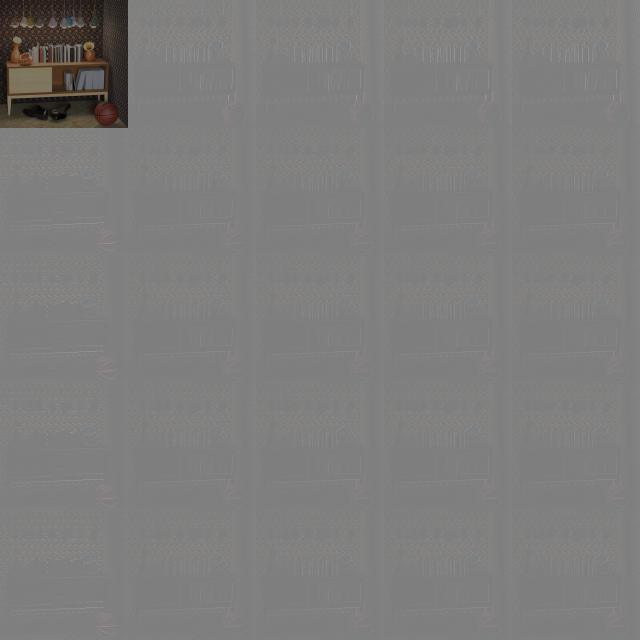}
		\caption{Reference image and prediction residuals resulting from the difference between a SAI and the average of two causal SAIs (top and left), Sideboard image from~\cite{Honauer_2016_ACoCV_ADatasetandEvaluationMethodologyforDepthEstimationon4dLightFields}. Displaying only the top-left $ 3 \times 3 $ SAIs.}
		\label{fig:4d-pred-example}
	\end{figure}

	As mentioned before, the focus of this work is the development of new native 4D frameworks, \textit{i.e.}, 4D prediction modes for MRP, to better exploit the pixel redundancies in the 4D space of LF representation $ L(t, s, v, u) $ described in Section~\ref{sec:4d-lf}. Such redundancies are evidenced in Figure~\ref{fig:4d-lf}, as discussed in the same section and confirmed by Figure~\ref{fig:4d-pred-example}, which shows the result of a zeroth-order angular prediction (\textit{i.e.}, prediction residue), obtained by subtracting the average of the top and left neighbouring SAIs from the current one, the top, and left ones. This simple experiment demonstrates that the energy of prediction residuals is quite low, indicating that more advanced prediction modes, which fully exploit the 4D space, are expected to achieve significant improvement in coding efficiency.
	
	\subsection{4D-MRP}
	
\begin{figure}
		\centering
		\resizebox{0.9\columnwidth}{!}{\input{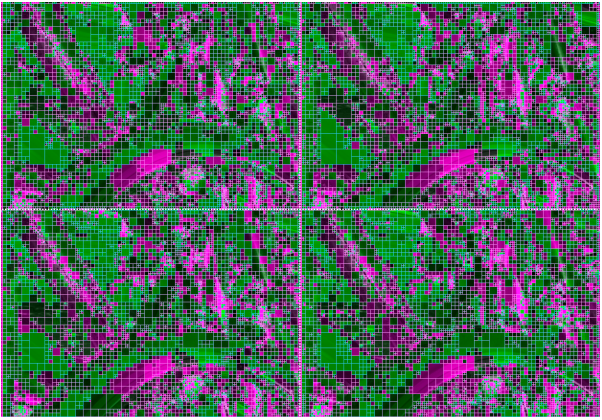}}
		\caption{Example of MRP class selection in neighbouring SAI, using Bikes image from~\cite{Rerabek_2016_8ICoQoMEQ}, where each class is represented by a different colour.}
		\label{fig:class-selection}
	\end{figure}

	The four-dimensional Minimum Rate Predictor (4D-MRP), stands on two main pillars: four dimensional prediction and partition of the LF into 4D blocks for the selection of the class. The motivation for using 4D blocks is supported by the fact that, in general, the same class is selected for co-located blocks in neighbouring SAIs. Figure~\ref{fig:class-selection} displays this behaviour, where each class is represented by a different colour. It can be seen that in each SAI, separated by white dashed lines, the same structures (\textit{e.g.} the wheel of the bicycle) presents the same colours in roughly the same areas. This means that the same class was selected for this area in the 4 SAIs. This information indicates that bitrate savings can be achieved in the signalling cost incurred by the quadtree. By using a 4D block instead of a 2D one, signalling each 2D block partitioning in each SAI is avoided.
	
	Therefore, in the novel prediction mode proposed in the 4D-MRP, causal neighbouring SAIs are used to extend the number of reference pixels in the MRP prediction. Since MRP is pixel-based, not all SAI pixels might be available for the current pixel, thus the need to use a causal vicinity. In this context, causality is defined by the encoding order of the SAIs, \textit{i.e.}, causal SAIs are the ones that were already encoded. The causal vicinity defines the SAIs that can be used as references, which is represented in Figure 6, where four causal SAI neighbours are used, namely: left ($ l $), left diagonal ($ tl $), top ($ t $), and right diagonal ($ tr $). The grey line divides the causal pixels. Equation~\ref{eq:mrp-pred} shows how the prediction is computed for pixel $ p_0 $, as a linear combination of the neighbouring pixels: 
	\begin{figure}[t]
		\centering
		\resizebox{\linewidth}{!}{\begin{tikzpicture}[anchor=south west, inner sep=0pt, scale=1]
	\centering
	\tikzstyle{pk_style}=[anchor=center, font=\normalsize, text centered]
	\def\radius{0.25}
	\def\cdist{(\radius * 2 + 0.1}
	
	\draw[step=3cm, very thick, dashed] (2.5,2.5) grid (12.5,9.5);
	
	\draw[step=3cm, line width=3.5pt] (6,3) grid (9,6);

	\begin{scope}[xshift=7.73cm, yshift=4.315cm]
		\begin{scope}[densely dotted, xshift=-2 * \cdist cm, yshift = -2 * \cdist cm, scale=0.7, pk_style, every node/.style={scale=0.6}]
			\foreach \x in {-3,...,3}
				{
				\foreach \y in {-2,...,4}
					{
					\draw(\x * \cdist, \y * \cdist) circle (\radius cm);
					}
				}
		\end{scope}
		
		\begin{scope}[xshift=-2 * \cdist cm, yshift = -2 * \cdist cm, scale=0.7, pk_style, every node/.style={scale=0.6}]
			\draw[very thick] (0, 0) circle (\radius cm) node {$p_0$};
			\draw (-\cdist, 0) circle (\radius cm) node {$p_1$};
			\draw (-2 * \cdist, 0) circle (\radius cm) node {$p_3$};
			\draw (-3 * \cdist, 0) circle (\radius cm) node {$p_7$};
			\draw (2 * \cdist, \cdist) circle (\radius cm) node {$p_{12}$};
			\draw (\cdist, \cdist) circle (\radius cm) node {$p_6$};
			\draw (0, \cdist) circle (\radius cm) node {$p_2$};
			\draw (-\cdist, \cdist) circle (\radius cm) node {$p_4$};
			\draw (-2 * \cdist, \cdist) circle (\radius cm) node {$p_8$};

			\draw (\cdist, 2 * \cdist) circle (\radius cm) node {$p_{11}$};
			\draw (0, 2 * \cdist) circle (\radius cm) node {$p_5$};
			\draw (-\cdist, 2 * \cdist) circle (\radius cm) node {$p_9$};

			\draw (0, 3 * \cdist) circle (\radius cm) node {$p_{10}$};

			\draw[gray, solid, thick] (-10.5 * \cdist,-0.5 * \cdist)--(-0.5 * \cdist,-0.5 * \cdist);
			\draw[gray, solid, thick] (-0.5 * \cdist,-0.5 * \cdist)--(-0.5 * \cdist,0.5 * \cdist);
			\draw[gray, solid, thick] (-0.5 * \cdist,0.5 * \cdist)--(3.5 * \cdist,0.5 * \cdist);
		\end{scope}
	\end{scope}
	
	\begin{scope}[xshift=7.73cm - 3cm, yshift=4.315cm]
		\begin{scope}[xshift=-2 * \cdist cm, yshift = -2 * \cdist cm, scale=0.7, pk_style, every node/.style={scale=0.6}]
			\draw[very thick] (0, 0) circle (\radius cm) node {$p_1$};
			\draw (-\cdist, 0) circle  (\radius cm) node {$p_2$};
			\draw (0, \cdist) circle (\radius cm)  node {$p_3$};
			\draw (\cdist, 0) circle (\radius cm)  node {$p_4$};
			\draw (-2 * \cdist, 0) circle (\radius cm)  node {$p_5$};
			\draw (-\cdist, \cdist) circle (\radius cm)  node {$p_6$};
			\draw (0, 2 * \cdist) circle (\radius cm)  node {$p_7$};
			\draw (\cdist, \cdist) circle (\radius cm)  node {$p_8$};
			\draw (2 * \cdist, 0) circle (\radius cm)  node {$p_{9}$};
			
			\draw[font=\Huge] (-4 * \cdist, 4 * \cdist) node {$ l $};
		\end{scope}

		\begin{scope}[densely dotted, xshift=-2 * \cdist cm, yshift = -2 * \cdist cm, scale=0.7, pk_style, every node/.style={scale=0.6}]
			\foreach \x in {-3,...,3}
				{
				\foreach \y in {-2,...,4}
					{
					\draw(\x * \cdist, \y * \cdist) circle (\radius cm);
					}
				}
		\end{scope}
	\end{scope}
	
	\begin{scope}[xshift=7.73cm - 3cm, yshift=4.315cm + 3cm]
		\begin{scope}[xshift=-2 * \cdist cm, yshift = -2 * \cdist cm, scale=0.7, pk_style, every node/.style={scale=0.6}]
			\draw[very thick] (0, 0) circle (\radius cm) node {$p_1$};
			\draw (-\cdist, 0) circle  (\radius cm) node {$p_2$};
			\draw (0, \cdist) circle (\radius cm)  node {$p_3$};
			\draw (\cdist, 0) circle (\radius cm)  node {$p_4$};
			\draw (0, -\cdist) circle (\radius cm)  node {$p_5$};
			\draw (-2 * \cdist, 0) circle (\radius cm)  node {$p_6$};
			\draw (-\cdist, \cdist) circle (\radius cm)  node {$p_7$};
			\draw (0, 2 * \cdist) circle (\radius cm)  node {$p_8$};
			\draw (\cdist, \cdist) circle (\radius cm)  node {$p_9$};
			\draw (2 * \cdist, 0) circle (\radius cm)  node {$p_{10}$};
			\draw (\cdist, -\cdist) circle (\radius cm)  node {$p_{11}$};
			\draw (0, -2 * \cdist) circle (\radius cm)  node {$p_{12}$};
			\draw (-\cdist, -\cdist) circle (\radius cm)  node {$p_{13}$};
			
			\draw[font=\Huge] (-3 * \cdist, 5.2 * \cdist) node {$ tl $};
		\end{scope}

		\begin{scope}[densely dotted, xshift=-2 * \cdist cm, yshift = -2 * \cdist cm, scale=0.7, pk_style, every node/.style={scale=0.6}]
			\foreach \x in {-3,...,3}
				{
				\foreach \y in {-2,...,4}
					{
					\draw(\x * \cdist, \y * \cdist) circle (\radius cm);
					}
				}
		\end{scope}
	\end{scope}
	
	\begin{scope}[xshift=7.73cm, yshift=4.315cm + 3cm]
		\begin{scope}[xshift=-2 * \cdist cm, yshift = -2 * \cdist cm, scale=0.7, pk_style, every node/.style={scale=0.6}]
			\draw[very thick] (0, 0) circle (\radius cm) node {$p_1$};
			\draw (-\cdist, 0) circle  (\radius cm) node {$p_2$};
			\draw (0, \cdist) circle (\radius cm)  node {$p_3$};
			\draw (\cdist, 0) circle (\radius cm)  node {$p_4$};
			\draw (0, -\cdist) circle (\radius cm)  node {$p_5$};
			\draw (-2 * \cdist, 0) circle (\radius cm)  node {$p_6$};
			\draw (-\cdist, \cdist) circle (\radius cm)  node {$p_7$};
			\draw (0, 2 * \cdist) circle (\radius cm)  node {$p_8$};
			\draw (\cdist, \cdist) circle (\radius cm)  node {$p_9$};
			\draw (2 * \cdist, 0) circle (\radius cm)  node {$p_{10}$};
			\draw (\cdist, -\cdist) circle (\radius cm)  node {$p_{11}$};
			\draw (0, -2 * \cdist) circle (\radius cm)  node {$p_{12}$};
			\draw (-\cdist, -\cdist) circle (\radius cm)  node {$p_{13}$};
			
			\draw[font=\Huge] (-3 * \cdist, 5.2 * \cdist) node {$ t $};
		\end{scope}

		\begin{scope}[densely dotted, xshift=-2 * \cdist cm, yshift = -2 * \cdist cm, scale=0.7, pk_style, every node/.style={scale=0.6}]
			\foreach \x in {-3,...,3}
				{
				\foreach \y in {-2,...,4}
					{
					\draw(\x * \cdist, \y * \cdist) circle (\radius cm);
					}
				}
		\end{scope}
	\end{scope}

	\begin{scope}[xshift=7.73cm + 3cm, yshift=4.315cm + 3cm]
		\begin{scope}[xshift=-2 * \cdist cm, yshift = -2 * \cdist cm, scale=0.7, pk_style, every node/.style={scale=0.6}]
			\draw[very thick] (0, 0) circle (\radius cm) node {$p_1$};
			\draw (-\cdist, 0) circle  (\radius cm) node {$p_2$};
			\draw (0, \cdist) circle (\radius cm)  node {$p_3$};
			\draw (\cdist, 0) circle (\radius cm)  node {$p_4$};
			\draw (0, -\cdist) circle (\radius cm)  node {$p_5$};
			\draw (-2 * \cdist, 0) circle (\radius cm)  node {$p_6$};
			\draw (-\cdist, \cdist) circle (\radius cm)  node {$p_7$};
			\draw (0, 2 * \cdist) circle (\radius cm)  node {$p_8$};
			\draw (\cdist, \cdist) circle (\radius cm)  node {$p_9$};
			\draw (2 * \cdist, 0) circle (\radius cm)  node {$p_{10}$};
			\draw (\cdist, -\cdist) circle (\radius cm)  node {$p_{11}$};
			\draw (0, -2 * \cdist) circle (\radius cm)  node {$p_{12}$};
			\draw (-\cdist, -\cdist) circle (\radius cm)  node {$p_{13}$};
			
			\draw[font=\Huge] (-3 * \cdist, 5.2 * \cdist) node {$ tr $};
		\end{scope}

		\begin{scope}[densely dotted, xshift=-2 * \cdist cm, yshift = -2 * \cdist cm, scale=0.7, pk_style, every node/.style={scale=0.6}]
			\foreach \x in {-3,...,3}
				{
				\foreach \y in {-2,...,4}
					{
					\draw(\x * \cdist, \y * \cdist) circle (\radius cm);
					}
				}
		\end{scope}
	\end{scope}
	
	\draw[step=3cm, very thick] (3,3) grid (12,9);
	\draw[red, step=3cm, very thick, dashed] (3,3) grid (6,6);
	\draw[green, step=3cm, very thick, dashed] (3,6) grid (6,9);
	\draw[cyan, step=3cm, very thick, dashed] (6,6) grid (9,9);
	\draw[yellow, step=3cm, very thick, dashed] (9,6) grid (12,9);
	
	\node(start) at (2.2 cm, 7.8 cm) {};
	\draw[latex-latex, very thick] (start) node[below = 0.1cm] {$t$} -- ++ (0, 2 cm) -- ++ (2 cm, 0) node[right = 0.1cm] {$s$};

	\node(start) at (9.2cm, 5 cm) {};
	\draw[latex-latex, very thick] (start) node[below = 0.1cm] {$v$} -- ++ (0, 0.8 cm) -- ++ (0.8 cm, 0) node[right = 0.1cm] {$u$};
\end{tikzpicture}}
		\caption{Representation of 4D prediction in MRP, where pixels are illustrated by the circles, the SAIs are symbolised by the dashed squares (labelled: left ($ l $), top-left ($ tl $), top ($ t $), and top-right ($ tr $)), and the colours are used to highlight the different reference SAIs.}
		\label{fig:mrp-4d-pred}
	\end{figure}
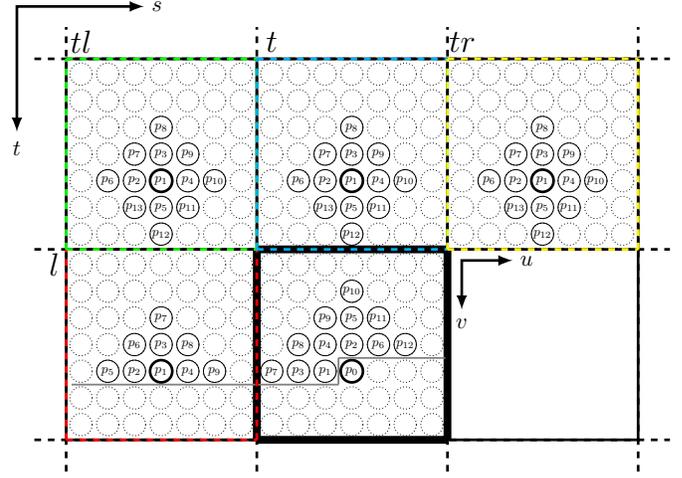
	\begin{equation}\label{eq:mrp-pred}
		\begin{split}
			\hat{s}(0) &= \sum\limits_{k = 1}^{K_c}a_{m}(c, k) \cdot s(c, k) + \sum\limits_{k = 1}^{K_{l}}a_{m}(l, k) \cdot s(l, k) \\
			&+ \sum\limits_{k = 1}^{K_{tl}}a_{m}(tl, k) \cdot s(tl,  k) + \sum\limits_{k = 1}^{K_{t}}a_{m}(t, k) \cdot s(t, k) \\
			&+ \sum\limits_{k = 1}^{K_{tr}}a_{m}(tr, k) \cdot s(tr, k),
		\end{split}
	\end{equation}
	where $ a_{m}(i, k) $, $ i = \{ c, l, tl, t, tr \} $ represent the prediction model weights associated with pixels in the current SAI (c) and the four causal neighbouring references, respectively. For the $ m \mhyphen th $ class, $ s(i, p_k) $, $ i = \{ c, l, tl, t, tr \}$ are the pixel values in the current and reference SAIs at position $ k $ of the prediction support region, and $ K_i $, $ i = \{c, l, tl, t, tr\} $ are the sizes of each prediction support region. The prediction support regions need not be the same size or shape.
	
	The context $ C $ associated with the $ p_0 $ pixel prediction error is calculated as the sum of the reference pixels prediction errors, $ e_{i,k} $, weighted by the inverse of the distances to these pixels, given by:
	\begin{equation}\label{eq:mrp-context}
		\begin{split}
			C &= \sum\limits_{k = 1}^{K_c} \frac{1}{\delta_{k}^{c}} e_c + \sum\limits_{k = 1}^{K_l} \frac{1}{\delta_{k}^{l}} e_l + \sum\limits_{k = 1}^{K_{tl}} \frac{1}{\delta_{k}^{tl}} e_{tl} \\
			&+ \sum\limits_{k = 1}^{K_t} \frac{1}{\delta_{k}^{t}} e_t + \sum\limits_{k = 1}^{K_{tr}} \frac{1}{\delta_{k}^{tr}} e_{tr},
		\end{split}
	\end{equation}
	where $ e_i = \left|s(i, k) - \hat{s}(i, k)\right| $ and $ \delta_{k}^{i},\ i = \{c, l, tl, t, tr\} $ are weighting factors proportional to the Euclidean distance between the current pixel ($ p_0 $) and their references ($ p_k $). The $ \delta_{k} $ weight is defined as $ \delta_{k} = \frac{\sqrt{d_v(k)^2 + d_u(k)^2 + d^2}}{64} $, where $ d $ represents the distance between SAIs and is set to 1 for all references except the current one, for which it is 0. The resulting $ C $ undergoes a quantisation step, using the thresholds described in Section~\ref{ssec:mrp-algorithm}, and is used to select the adaptive conditional probability models.
	
	In MRP, a set of prediction coefficients, $ a_{m}(i, k) $, are determined for each class independently. Then each pixel of the current SAI is assigned a class corresponding to  the predictor that minimises Equation~\ref{eq:mrp-cost}. By using 4D blocks, the MRP quadtree partition is expanded into a hexadecatree, where each block is divided in all four dimensions simultaneously, resulting in 16 sub-blocks in the 4D space. Let a block be represented by $ \mathcal{L}_B(t, s, v, u) $, with $ t \in \{\gamma \in \mathbb{N} \,| \,0\leq \gamma < T\} $, $ s \in \{\gamma \in \mathbb{N} \,| \,0\leq \gamma < S\} $, $ v \in \{\gamma \in \mathbb{N} \,| \,0\leq \gamma < V\} $, and $ u \in \{\gamma \in \mathbb{N} \,| \,0\leq \gamma < U\} $, in the block coordinates, then the hexadecatree partition can be represented by Equation~\ref{eq:hex-part}.
	\begin{align}\label{eq:hex-part}
		\begin{array}{l}
			P_{\mathrm{Hex}}(\mathcal{L}_{B}(t, s, v, u)) = \\ \hspace{0.75cm}\left\{ \mathcal{L}_{B_i}(t_i, s_i, v_i, u_i), \forall i \in \{0, \ldots, 15\} \right\},
		\end{array}
		\nonumber\\[5pt]
		\mbox{ with }
		\left\{
		\begin{array}{l}
			t_i = a + T\times (1 + (-1)^{\lfloor i/8 \rfloor + 1}) / 4, \\
			s_i = b + S\times (1 + (-1)^{\lfloor i/4 \rfloor + 1}) / 4, \\
			v_i = c + V\times (1 + (-1)^{\lfloor i/2 \rfloor + 1}) / 4, \\
			u_i = d + U\times (1 + (-1)^{i+1})/4, \\ \\
			\forall a \in \{\gamma \in \mathbb{N} \,| \,0\leq \gamma < T / 2\}, \\
			\forall b \in \{\gamma \in \mathbb{N} \,| \,0\leq \gamma < S / 2\}, \\
			\forall c \in \{\gamma \in \mathbb{N} \,| \,0\leq \gamma < V / 2\}, \\
			\forall d \in \{\gamma \in \mathbb{N} \,| \,0\leq \gamma < U / 2\} \\
		\end{array} \right.,
	\end{align}
	where $ T $, $ S $, $ V $ and $ U $ represent the dimensions of the 4D block.
	
\begin{figure}[t]
		\centering
		\resizebox{\linewidth}{!}{
\tikzstyle{line} = [draw, -latex']
\tikzstyle{tnode} = [draw, circle, node distance=0.5cm, minimum height=0.3cm]
    
\begin{tikzpicture}[node distance = 2cm, font=\scriptsize, auto]
    \node [tnode] (origin) {};
    \node [tnode, below of=origin, below=1cm, label=center:7] (1) {};
    \path [line] (origin) -- (1);
    
    \foreach \x in {2,3,...,8}
      {
		\pgfmathparse{int(\x-1)}
		\edef\ref{\pgfmathresult}
		
		\pgfmathparse{int(8-\x)}
		\edef\label{\pgfmathresult}
		
		\node [tnode, left of=\ref, label=center:\label] (\x) {};
		
		\path [line] (origin) -- (\x);
      }
      
	\foreach \x in {2,3,...,9}
      {
		\pgfmathparse{int(\x-1)}
		\edef\ref{\pgfmathresult}
		
		\pgfmathparse{int(6+\x)}
		\edef\label{\pgfmathresult}
		
		\node [tnode, right of=\ref, label=center:\label] (\x) {};
		
		\path [line] (origin) -- (\x);
      }

	\node [circle, node distance=0.5cm, minimum height=0.3cm, right of=2,white] (origin2) {};
	\node [tnode, below of=3, below=1cm, label=center:7] (1) {};
    \path [line] (origin2) -- (1);
    
	\foreach \x in {2,3,...,8}
      {
		\pgfmathparse{int(\x-1)}
		\edef\ref{\pgfmathresult}
		
		\pgfmathparse{int(8-\x)}
		\edef\label{\pgfmathresult}
		
		\node [tnode, left of=\ref, label=center:\label] (\x) {};
		
		\path [line] (origin2) -- (\x);
      }
      
	\foreach \x in {2,3,...,9}
      {
		\pgfmathparse{int(\x-1)}
		\edef\ref{\pgfmathresult}
		
		\pgfmathparse{int(6+\x)}
		\edef\label{\pgfmathresult}
		
		\node [tnode, right of=\ref, label=center:\label] (\x) {};
		
		\path [line] (origin2) -- (\x);
      }

\end{tikzpicture}}
		\caption{Hexadecatree partition in the 4D-MRP: tree structure partition example in which a top block is partitioned into 16 4D sub-blocks (numbered 0 to 15) and an equivalent partition of the sub-block 9.}
		\label{fig:hexadecatree-flow}
	\end{figure}
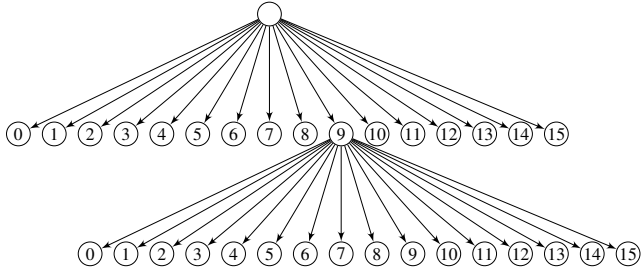

\begin{figure}[t]
		\centering
		\resizebox{0.9\linewidth}{!}{\input{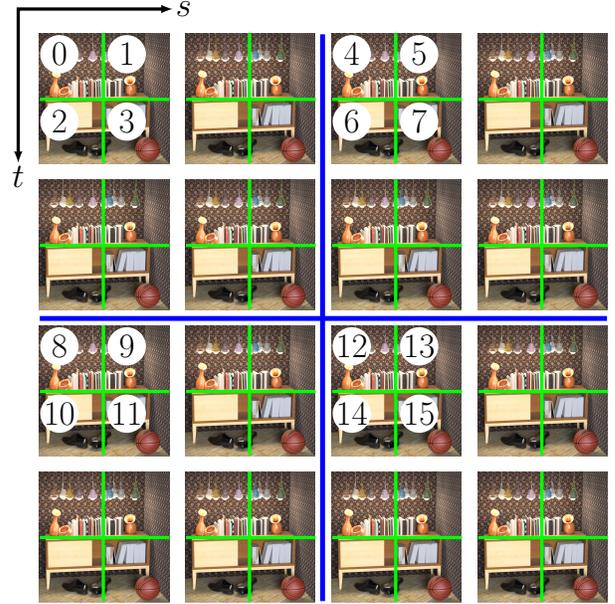}}
		\caption{Hexadecatree partition in the 4D-MRP: visual representation of the 4D block partition, where the blue lines represent the angular partition and the green lines the spatial partition. The partition index of each block is shown in white circles.}
		\label{fig:hexadecatree}
	\end{figure}

	Figure~\ref{fig:hexadecatree-flow} represents the tree structure partition and Figure~\ref{fig:hexadecatree} shows a visual representation of the 4D block partition. In Figure~\ref{fig:hexadecatree} the angular dimensions $ (t, s) $ are partitioned by the green lines and spatial dimensions $ (v, u) $ by the blue lines, the index of each 4D block is shown in white circles. As pointed out before, the hexadecatree partitioning is expected to result in a larger number of pixels being classified with lower signalling cost, thus increasing compression efficiency.
	
	The hexadecatree partition and block classification optimisation is performed at the \textit{Optimise Classification} stage, shown in Figure~\ref{fig:mrp-fluxogram}. This optimization aims at minimising both side information, represented by $ B_m $ in Equation~\ref{eq:mrp-cost}, and the prediction error cost. $ B_m $ can be further represented by:
	
	\begin{equation}
		B_m = B_{m_{\mathrm{flags}}} + B_{m_{\mathrm{class}}},
	\end{equation}
	where $ B_{m_{\mathrm{flags}}} $, given by Equation~\ref{eq:4dmrp-bmflags}, represents the cost of selecting the partition flags and $ B_{m_{\mathrm{class}}} $ the cost of selecting a class for each resulting block.
	\begin{equation}\label{eq:4dmrp-bmflags}
		B_{m_{\mathrm{flags}}} = 
		\begin{cases}
			-\log_2(P_{\mathrm{ctx}}), & \text{not partitioned} \\
			-\log_2(1 - P_{\mathrm{ctx}}), &\text{partitioned}
		\end{cases}
	\end{equation}
	In Equation~\ref{eq:4dmrp-bmflags}, $ P_{\mathrm{ctx}} $ represents an approximation of the probability of the blocks in a given context not being partitioned, given by:
	\begin{equation}\label{eq:prob-4dmrp}
		\begin{split}
			P_{\mathrm{ctx}} &= \\ &\underset{p}{\mathrm{argmin}} \left\{-\log_2(p) \times \#\mathit{ctx}_0 \right. \left.-\log_2(1 - p) \times \#\mathit{ctx}_1 \right\},\\&\text{with \qquad} p \in \{0.05, 0.2, 0.35, 0.5, 0.65, 0.8, 0.95\},
		\end{split}
	\end{equation}
	where $ \#\mathit{ctx}_0 $ and $ \#\mathit{ctx}_1 $ represent the number of occurrences of the $ 0 $ and $ 1 $ partition flags, respectively, in each context, \textit{i.e.} the number of times each block is either not-partitioned or partitioned in each given context. $ p $ represents the probability that minimises the cost of selecting a flag in each context. This probability is selected from a discrete set to minimise the cost of transmitting the real probability value, therefore only a single index needs to be sent to the decoder. The quantified probability calculation of Equation~\ref{eq:prob-4dmrp} follows that used in MRP, which was first introduced with the variable block size in~\cite{Matsuda_2007_SaCiJ_ALosslessCodingSchemeUsingAdaptivePredictorsandArithmeticCodeOptimizedforEachImage}.
	
	Finally, the context calculation is determined at each level by the number neighbouring blocks which were partitioned at said level:
	\begin{equation}
		\mathrm{C_{qtflags}}(\mathit{level}, b) = \sum_{b \in N} \mathrm{qtflags}(\mathit{level}, b),
	\end{equation}
	where $ b $ represents a block, $ \mathrm{qtflags} $ the partition flag for block $ b $ with the size given by $ \mathit{level} $, $ 0 $ and $ 1 $ for non-partitioned and partitioned blocks, respectively, and $ N $ represents the neighbouring blocks on the same level as $ b $. As 4D-MRP uses 4 levels for the block sizes and in the hexadecatree 5 neighbours are being considered, there are a total of 24 different contexts.
	
	In the \textit{Design Predictors} stage of the encoding process, fixed size blocks are used to determine the linear predictor coefficients of each class. These operations are affected by the block size, as is the coding performance, through the calculation of the coefficients $ a_{m_i}(k) $, which decrease with the larger number of pixels. Experiments showed that the $ 8 \times 8 \times 8 \times 8 $ pixels fixed block-size, analogous to what is used in MRP, was too large for the 4D space, which resulted in blocks of $ 8 ^ 4 = 4096 $ pixels total. The same experiments showed that blocks of $ 4 ^ 4 = 256 $ pixels were a good compromise in terms of coding ratio. Thus in this work, when 4D blocks are used in the \textit{Fixed Block-Size} part, the size is set to $ 4 \times 4 \times 4 \times 4 $. In the \textit{Variable Block-Size} part the top level block size is set to $ 32 \times 32 \times 32 \times 32 $ pixels.
	
	\subsection{Dual-Tree 4D-MRP (DT-4D-MRP)}\label{sec:dt-4d-mrp}
	
	\subsubsection{Dual-Tree Partitioning}
	
	This partitioning mode is intended to improve the problem created by hexadecatrees, which tend to use very fine partitioning with a high number of $ 2 \times 2 \times 2 \times 2 $ blocks (the smallest possible size), albeit still achieving better performance than MRP, as shall be seen in Section~\ref{sec:experimental-results}. As expected,  using smaller  block sizes requires more signalling bits to encode the partition tree. An analysis of the costs and benefits of the partitioning alternatives leads to the hypothesis that independently partitioning the spatial and angular dimensions could lead to further coding gains. This is supported by different sampling present on the spatial and angular dimensions at the acquisition.
	
	In this mode, the block partitioning algorithm was changed to decouple the angular from the spatial partitioning. Therefore, rather than partitioning in 16 blocks as in 4D-MRP, each 4D block can be partitioned either into four blocks in the angular dimensions $ (t, s) $, into four blocks in the spatial dimensions $ (v, u) $, or not partitioned at all. Let a block be represented by $ \mathcal{L}_B(t, s, v, u) $, with $ t \in \{\gamma \in \mathbb{N} \,| \,0\leq \gamma < T\} $, $ s \in \{\gamma \in \mathbb{N} \,| \,0\leq \gamma < S\} $, $ v \in \{\gamma \in \mathbb{N} \,| \,0\leq \gamma < V\} $, and $ u \in \{\gamma \in \mathbb{N} \,| \,0\leq \gamma < U\} $, in the block coordinates. Then the dual-tree partition can be defined by Equations~\ref{eq:dt-part-s} and \ref{eq:dt-part-a}, for the spatial and angular partitions respectively.
	\begin{align}\label{eq:dt-part-s}
		P_{\mathrm{DT_S}}(\mathcal{L}_{B}(t, s, v, u)) = \left\{ \mathcal{L}_{B_i}(t, s, v_i, u_i), \forall i \in \{0, \ldots, 3\} \right\}, \nonumber\\[5pt]
		\mbox{\quad with } \left\{
		\begin{array}{l}
			\forall t \in \{\gamma \in \mathbb{N} \,| \,0\leq \gamma < T\},\\
			\forall s \in \{\gamma \in \mathbb{N} \,| \,0\leq \gamma < S\},\\
			v = a + V\times (1 + (-1)^{\lfloor i / 2 \rfloor+1})/4,\\
			u = b + U\times (1 + (-1)^{i + 1}) / 4, \\ \\
			\forall a \in \{\gamma \in \mathbb{N} \,| \,0\leq \gamma < V / 2\},\\
			\forall b \in \{\gamma \in \mathbb{N} \,| \,0\leq \gamma < U / 2\}
		\end{array} \right. .
	\end{align}
	\begin{align}\label{eq:dt-part-a}
		P_{\mathrm{DT_A}}(\mathcal{L}_{B}(t, s, v, u)) = \left\{ \mathcal{L}_{B_i}(t_i, s_i, v, u), \forall i \in \{0, \ldots, 3\} \right\}, \nonumber\\[5pt]
		\mbox{\quad with } \left\{
		\begin{array}{l}
			t = a + T\times (1 + (-1)^{\lfloor i / 2 \rfloor + 1}) / 4,\\
			s = b + S\times (1 + (-1)^{i + 1}) / 4,\\
			\forall v \in \{\gamma \in \mathbb{N} \,| \,0\leq \gamma < V\},\\
			\forall u \in \{\gamma \in \mathbb{N} \,| \,0\leq \gamma < U\},\\ \\
			\forall a \in \{\gamma \in \mathbb{N} \,| \,0\leq \gamma < T / 2\},\\
			\forall b \in \{\gamma \in \mathbb{N} \,| \,0\leq \gamma < S / 2\}
		\end{array} \right. .
	\end{align}
	
\begin{figure}[t]
		\centering
		\resizebox{0.65\linewidth}{!}{
\tikzstyle{line} = [draw, -latex']
\tikzstyle{tnode} = [draw, circle, node distance=0.8cm, minimum height=0.5cm]
\tikzstyle{tnode-label} = [node distance=0.32cm, font=\tiny]
    
\begin{tikzpicture}[node distance = 2cm,, auto]
    \node [tnode] (origin) {};
	\node [tnode-label, right of=origin] {S};
    \node [tnode, below of=origin, node distance=1cm, left=0.15cm, label=center:1] (1) {};
	\node [tnode-label, right of=1] {N};
    
    \path [line] (origin) -- (1);
    
    \foreach \x in {2}
      {
		\pgfmathparse{int(\x-1)}
		\edef\ref{\pgfmathresult}
		
		\pgfmathparse{int(2-\x)}
		\edef\label{\pgfmathresult}
		
		\node [tnode, left of=\ref, label=center:\label] (\x) {};
		\node [tnode-label, right of=\x] {S};
		
		\path [line] (origin) -- (\x);
      }
      
	\foreach \x in {2,3}
      {
		\pgfmathparse{int(\x-1)}
		\edef\ref{\pgfmathresult}
		
		\pgfmathparse{int(0+\x)}
		\edef\label{\pgfmathresult}
		
		\node [tnode, right of=\ref, label=center:\label] (\x) {};
		
		\path [line] (origin) -- (\x);
      }
      
	\node [tnode-label, right of=2] {N};

	\node [circle, node distance=0.8cm, minimum height=0.5cm,white, right of=2] (origin2) {};
	\node [tnode-label, right of=origin2]  {A};
	\node [tnode, below of=3, node distance=1cm, left=0.15cm, label=center:1] (1) {};
	\node [tnode-label, right of=1] {A};
	
    \path [line] (origin2) -- (1) ;
    
	\foreach \x in {2}
      {
		\pgfmathparse{int(\x-1)}
		\edef\ref{\pgfmathresult}
		
		\pgfmathparse{int(2-\x)}
		\edef\label{\pgfmathresult}
		
		\node [tnode, left of=\ref, label=center:\label] (\x) {};
		
		\path [line] (origin2) -- (\x);
		
		\node [tnode-label, right of=\x] {S};
      }
      
	\foreach \x in {2,3}
      {
		\pgfmathparse{int(\x-1)}
		\edef\ref{\pgfmathresult}
		
		\pgfmathparse{int(0+\x)}
		\edef\label{\pgfmathresult}
		
		\node [tnode, right of=\ref, label=center:\label] (\x) {};
		
		\path [line] (origin2) -- (\x);
	
		\node [tnode-label, right of=\x] {N};
      }

\end{tikzpicture}}
		\caption{Dual quadtree partition in 4DMRP: tree structure partition example.}
		\label{fig:dualquadtree-flow}
	\end{figure}
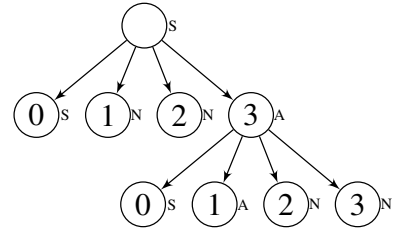

\begin{figure*}[t]
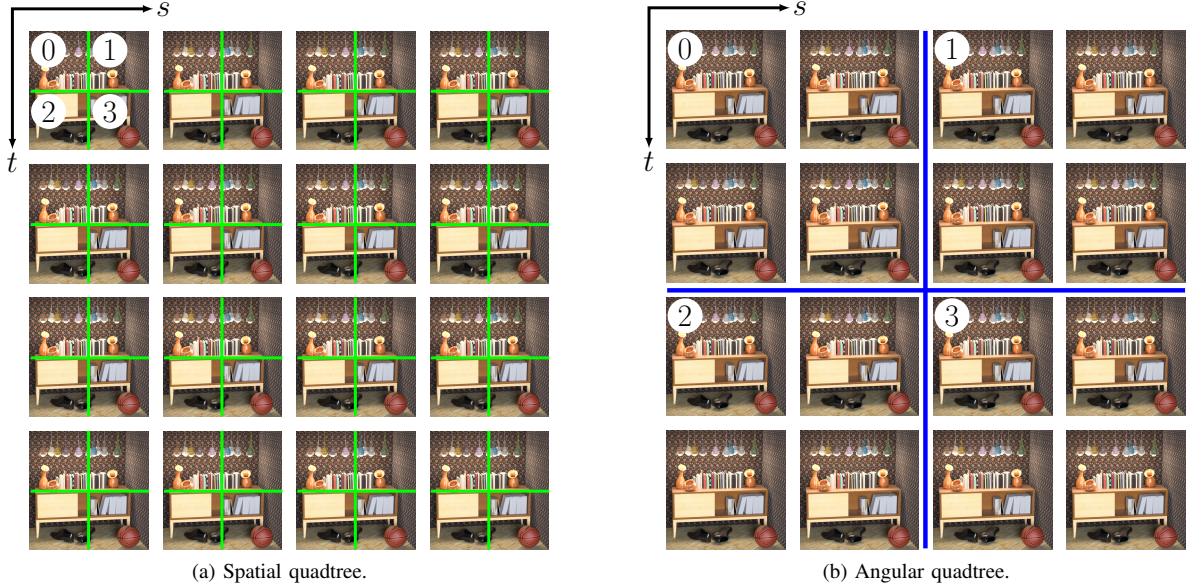

		\centering
		\subfloat[\label{fig:dualquadtree-spatial}Spatial quadtree.]{\resizebox{0.4\linewidth}{!}{\input{figures/spatial-quadtree.pgf}}}\hfil
		\subfloat[\label{fig:dualquadtree-angular}Angular quadtree.]{\resizebox{0.4\linewidth}{!}{\input{figures/angular-quadtree.pgf}}}
		\caption{Dual quadtree partition in 4DMRP, using Sideboard image from~\cite{Honauer_2016_ACoCV_ADatasetandEvaluationMethodologyforDepthEstimationon4dLightFields}. The partition index of each block is shown in white circles.}
		\label{fig:dualquadtree}
	\end{figure*}

	This structure of 4D blocks is represented in Figure~\ref{fig:dualquadtree-flow}, where `N' represents the non-partitioned blocks,`S' the spatial partition, and `A' the angular partition. Figures~\ref{fig:dualquadtree-spatial} and \ref{fig:dualquadtree-angular} show visual representations of the separate quadtrees for spatial and angular dimensions, respectively.
	
	\subsubsection{Arithmetic Coding of the Partition Tree Signalling}
	
	In this new tree structure the decisions are no longer binary but ternary, since there are three options for partitioning each block (\textit{i.e.}, spatial, angular, none). In DT-4D-MRP, the cost of selecting a given partition mode was redefined to reflect the new signalling structure:
	\begin{equation}
		B_{m_{\mathrm{flags}}} = 
		\begin{cases}
			-\log_2(P_{N}), & \text{if flag = `N'} \\
			-\log_2(P_{S | \bar{N}} \cdot P_{\bar{N}}), & \text{if flag = `S'} \\
			-\log_2((1 - P_{S | \bar{N}}) \cdot P_{\bar{N}}), & \text{if flag = `A'}
		\end{cases},
	\end{equation}
	where `N', `S', and `A' represent the partition flags of a block, indicating whether a block was not partitioned, partitioned in the spatial dimensions, or partitioned in the angular dimension, respectively, as shown in Figure~\ref{fig:dualquadtree-flow}. $ P_{N} $ represents the probability of the `N' flag, $ P_{\bar{N}} $ the complementary probability of $ P_{N} $, and $ P_{S} $ the probability of the 'S' flag given $ \bar{N} $, resulting from:
	\begin{equation}
		\begin{split}
			&P_{N} = \\& \underset{p}{\mathrm{argmin}} \left\{-\log_2(p) \times \mathit{ctx}_N \right.  \left. -\log_2(1 - p) \times (\mathit{ctx}_S + \mathit{ctx}_A) \right\}, \\
			&P_{S | \bar{N}} = \underset{p}{\mathrm{argmin}} \left\{-\log_2(p) \times \mathit{ctx}_S \right. \left.-\log_2(1 - p) \times \mathit{ctx}_A \right\},
		\end{split}
	\end{equation}
	where $ \mathit{ctx}_i $ represents $ \mathrm{context}_{DT}(i) $, given by Equation~\ref{eq:dt-4dmrp-context} and $ p \in \{0.05, 0.2, 0.35, 0.5, 0.65, 0.8, 0.95\} $.
	
	For the DT-4D-MRP the number of possible contexts was reduced, and the computation of the context was replaced by the number of times each flag (`N', `S', or `A') is chosen in the tree:
	\begin{equation}\label{eq:dt-4dmrp-context}
		\mathrm{C_{qtflags}}_{DT}(F) = \sum_{b \in \mathrm{tree}} \delta_{\mathrm{flag}(b), F},
	\end{equation}
	where $ F $ represents the flag for which the context is being determined, $ b $ represents each block in the tree, $ \mathrm{flag}(b) $ is the partition flag of each block $ b $ and $ \delta $ is the Kronecker delta.
	
	These operations are represented by Algorithm~\ref{alg:flag-cost-dt-4d-mrp}. As in the 4D-MRP case, a table of probabilities (\textit{treeprob}) is used to facilitate the encoding of the partition flags probability, as the index is more efficiently encoded than the actual probability value. The optimal index $ o\_index $ corresponds to the probability $ p $ that minimises the expression in Line 6 and Line 18 of the algorithm. The cost of each flag is represented by the variable: $ \text{cost} = -log(P) $. This calculation is performed in two steps, initially the cost of the `N' flag is determined, by grouping the probability of the remaining flags, `S' and `A'. In the second step, only `S' and `A' are taken into account, and their relative probabilities are determined. For the actual cost, these probabilities are normalised by multiplying them by a factor of $ 1 - p_N $, such that $ \sum_{j = N, S, A} P_j = 1 $. These values are used to set the probability models for the arithmetic coding of the tree partition, and the previously mentioned index is explicitly transmitted in the bitstream. Each flag cost is used in the optimisation of the partition tree. On the decoder side, the partition flags are decoded using the transmitted probability indices, without the need of recalculating the contexts of Equation~\ref{eq:dt-4dmrp-context}.
	
	\begin{algorithm}
		\caption{Calculation of the partition flags cost.}
		\label{alg:flag-cost-dt-4d-mrp}
		\begin{algorithmic}[1]
			\State cost = inf
			\State treeprob $ = [0.05, 0.2, 0.35, 0.5, 0.65, 0.8, 0.95] $
			\State \#Determine cost of symbol `N'
			\For{$i = 0$ to $6$}
			\State $ p = $ treeprob[$i$]
			\State $ c = -\log(p) \cdot $ context[`N'] $ - \log(1 - p) \cdot ($context[`S'] $ + $ context[`A']$) $
			\If{ $ c < $ cost}
			\State cost $ = c $
			\State o\_index $ = i $
			\EndIf
			\EndFor
			\State $ p_N = $ treeprob$[\text{o\_index}]$
			\State flagcost[`N'] $ = - \log_2(p_N)$
			\State cost = inf
			\State \#Determine cost of symbols `S' and `A'
			\For{$i = 0$ to $6$}
			\State $ p = $ treeprob[$i$]
			\State $ c = -\log(p) \cdot $ context[`S'] $ - \log(1 - p) \cdot $ context[`A']
			\If{ $ c < $ cost}
			\State cost $ = c $
			\State o\_index $ = i $
			\EndIf
			\EndFor
			\State $ p = $ treeprob$[\text{o\_index}]$
			\State flagcost[`S'] $ = -\log_2(p \cdot (1 - p_N)) $
			\State flagcost[`A'] $ = -\log_2((1 - p) \cdot (1 - p_N)) $
		\end{algorithmic}
	\end{algorithm}
	
	The remaining steps in the arithmetic coding of the partition flags and class selection are kept unchanged. Essentially, starting with the first block, the partition flags are encoded until a block is no longer partitioned, then the selected class is encoded in the bitstream.
	
	\subsection{M-MRP}
	
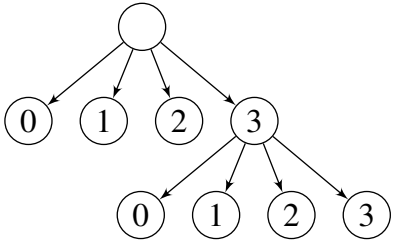
\begin{figure}[t]
		\centering
		\resizebox{0.65\linewidth}{!}{
\tikzstyle{line} = [draw, -latex']
\tikzstyle{tnode} = [draw, circle, node distance=0.8cm, minimum height=0.5cm]
\tikzstyle{tnode-label} = [node distance=0.32cm, font=\tiny]
    
\begin{tikzpicture}[node distance = 2cm,, auto]
    \node [tnode] (origin) {};
    \node [tnode, below of=origin, node distance=1cm, left=0.15cm, label=center:1] (1) {};
    
    \path [line] (origin) -- (1);
    
    \foreach \x in {2}
      {
		\pgfmathparse{int(\x-1)}
		\edef\ref{\pgfmathresult}
		
		\pgfmathparse{int(2-\x)}
		\edef\label{\pgfmathresult}
		
		\node [tnode, left of=\ref, label=center:\label] (\x) {};
		
		\path [line] (origin) -- (\x);
      }
      
	\foreach \x in {2,3}
      {
		\pgfmathparse{int(\x-1)}
		\edef\ref{\pgfmathresult}
		
		\pgfmathparse{int(0+\x)}
		\edef\label{\pgfmathresult}
		
		\node [tnode, right of=\ref, label=center:\label] (\x) {};
		
		\path [line] (origin) -- (\x);
      }
      

	\node [circle, node distance=0.8cm, minimum height=0.5cm,white, right of=2] (origin2) {};
	\node [tnode, below of=3, node distance=1cm, left=0.15cm, label=center:1] (1) {};
	
    \path [line] (origin2) -- (1) ;
    
	\foreach \x in {2}
      {
		\pgfmathparse{int(\x-1)}
		\edef\ref{\pgfmathresult}
		
		\pgfmathparse{int(2-\x)}
		\edef\label{\pgfmathresult}
		
		\node [tnode, left of=\ref, label=center:\label] (\x) {};
		
		\path [line] (origin2) -- (\x);
		
      }
      
	\foreach \x in {2,3}
      {
		\pgfmathparse{int(\x-1)}
		\edef\ref{\pgfmathresult}
		
		\pgfmathparse{int(0+\x)}
		\edef\label{\pgfmathresult}
		
		\node [tnode, right of=\ref, label=center:\label] (\x) {};
		
		\path [line] (origin2) -- (\x);
	
      }

\end{tikzpicture}}
		\caption{Two dimension quadtree partition in M-MRP: tree structure partition example.}
		\label{fig:2d-quadtree-flow}
	\end{figure}

\begin{figure}[t]
		\centering
		\resizebox{0.85\linewidth}{!}{\input{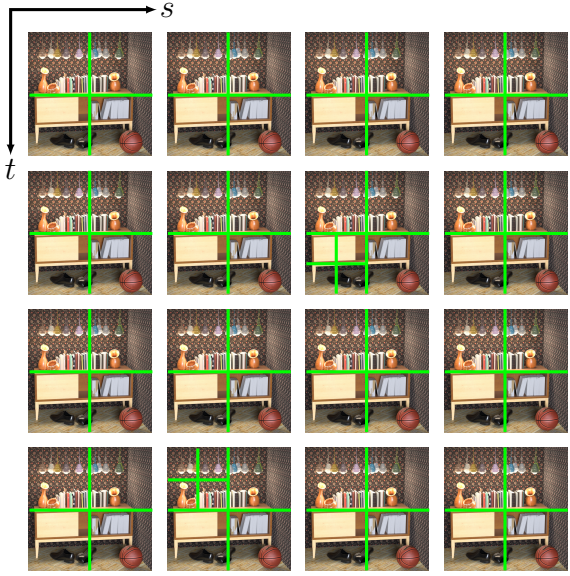}}
		\caption{Two dimension quadtree partition in M-MRP, using Sideboard image from~\cite{Honauer_2016_ACoCV_ADatasetandEvaluationMethodologyforDepthEstimationon4dLightFields}: visual representation of the quadtree using different levels.}
		\label{fig:2d-quadtree}
	\end{figure}

	Another alternative to deal with the large overhead of hexadecatree partitioning signalling is to abandon 4D partitioning and perform only 2D partitions. This is done by treating the LF as a single 2D image for the partition, which results in a better spatial partition, but forsakes the potential benefits of the 4D partition presented in 4D-MRP. The multiple reference MRP (M-MRP), which, as mentioned before, was described in~\cite{Santos_2019_DCC}, combines the 4D prediction described in the Section~\ref{sec:4d-mrp} (represented in Figure~\ref{fig:mrp-4d-pred}) with the conventional 2D quadtree partition, represented in Figure~\ref{fig:2d-quadtree-flow} and Figure~\ref{fig:2d-quadtree}. Let a 2D block be represented by $ \mathcal{L}_B(v, u) $, with $ v \in \{\gamma \in \mathbb{N} \,| \,0\leq \gamma < V\} $, and $ u \in \{\gamma \in \mathbb{N} \,| \,0\leq \gamma < U\} $, in the block coordinates, then the M-MRP partition can be represented by:
	\begin{align}
		P_{\mathrm{DT_S}}(\mathcal{L}_{B}(v, u)) = \left\{ \mathcal{L}_{B_i}(v_i, u_i), \forall i \in \{0, \ldots, 3\} \right\}, \nonumber\\[5pt]
		\mbox{\quad with } \left\{
		\begin{array}{l}
			v = a + V\times (1 + (-1)^{\lfloor i / 2 \rfloor+1})/4,\\
			u = b + U\times (1 + (-1)^{i + 1}) / 4, \\ \\
			\forall a \in \{\gamma \in \mathbb{N} \,| \,0\leq \gamma < V / 2\},\\
			\forall b \in \{\gamma \in \mathbb{N} \,| \,0\leq \gamma < U / 2\}
		\end{array} \right. .
	\end{align}
	
	Regarding the partition structure, as this is similar to 4D-MRP, in particular, and to MRP, in general, no further modifications to the signalling structure are needed.
	
	\section{Experimental Evaluation} \label{sec:experimental-results}
	
\begin{figure*}[t]
		\centering
		\subfloat[Bikes, from the EPFL dataset~\cite{Rerabek_2016_8ICoQoMEQ}.]{\includegraphics[height=4cm]{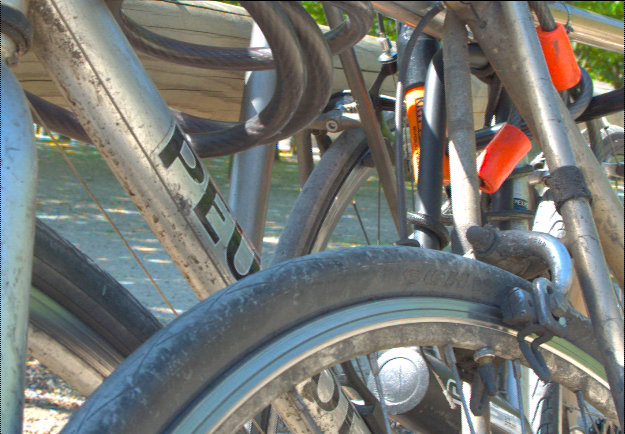}}
		\hfil
		\subfloat[Sideboard, from the HCI dataset~\cite{Honauer_2016_ACoCV_ADatasetandEvaluationMethodologyforDepthEstimationon4dLightFields}.]{\includegraphics[height=4cm]{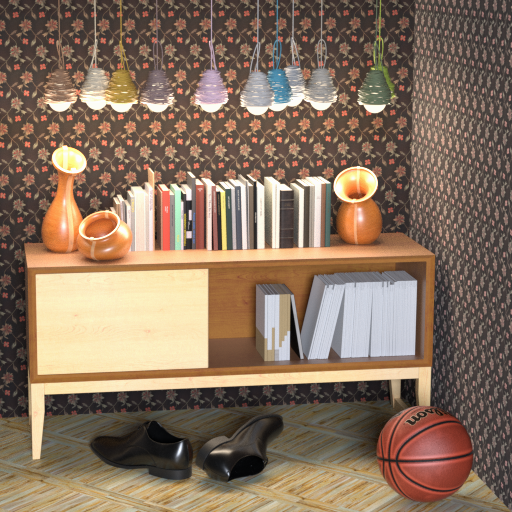}}
		\hfil
		\subfloat[Img1, from the SKINL2 dataset~\cite{Faria_2019_24AICotIEiMaBSE_LightFieldImageDatasetofSkinLesions}.]{\includegraphics[height=4cm]{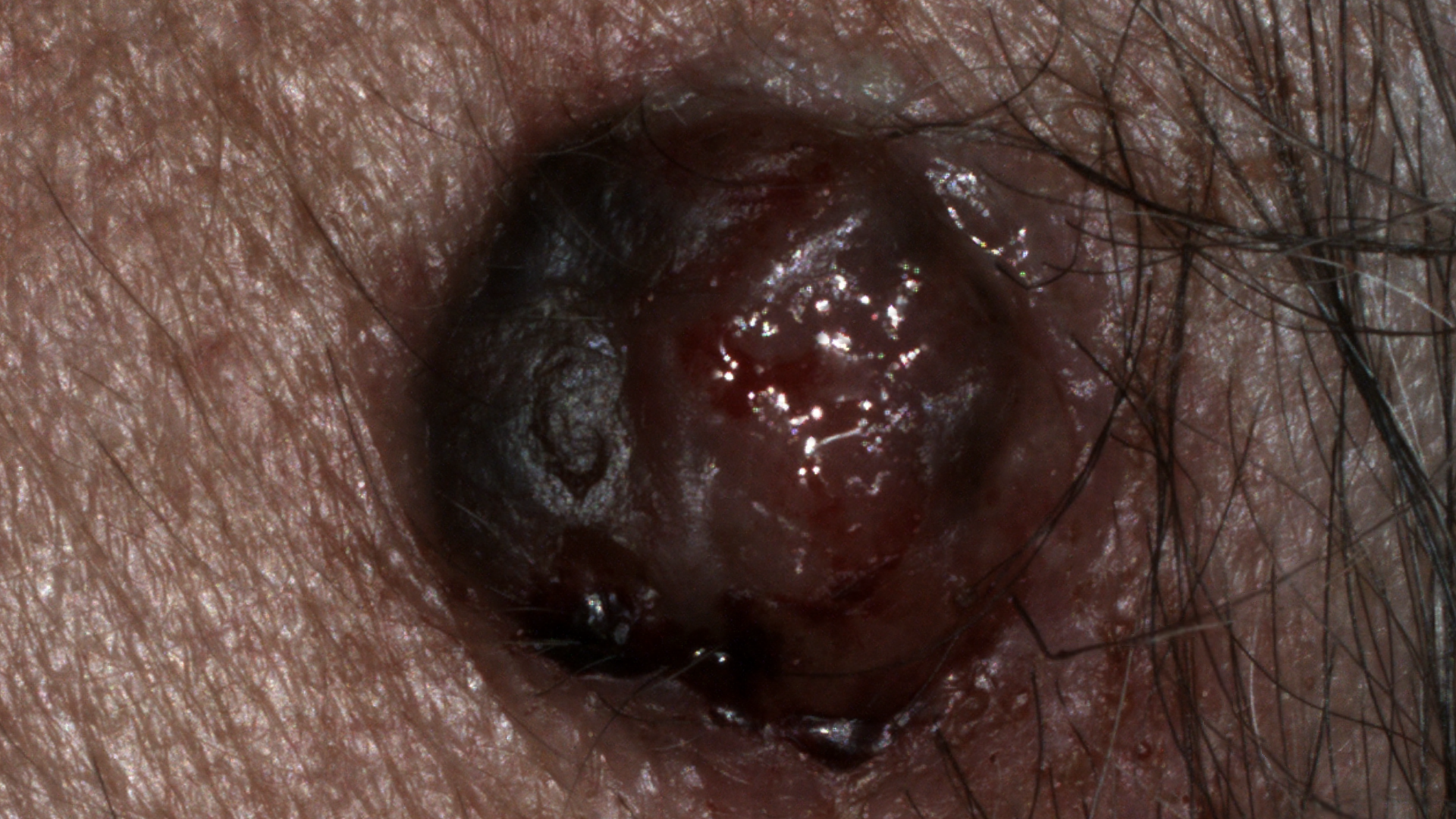}}
		\caption{Sub-aperture image from one image of each of the used datasets.}
		\label{fig:dataset-examples}
	\end{figure*}

	The experimental evaluation follows the common test conditions of JPEG-Pleno (JPEG-Pleno CTC)~\cite{JPEG-CTC}, with minor changes to adapt to lossless compression. The experiments use two of the JPEG-Pleno CTC datasets, namely the Lenslet Lytro Illum Camera dataset from EPFL~\cite{Rerabek_2016_8ICoQoMEQ} and the Synthetic HCI HDCA dataset from HCI~\cite{Honauer_2016_ACoCV_ADatasetandEvaluationMethodologyforDepthEstimationon4dLightFields}. Additionally, to expand the range of these experiments a dataset of skin lesions, Skin Lesion Light-fields (SKINL2)~\cite{Faria_2019_24AICotIEiMaBSE_LightFieldImageDatasetofSkinLesions}, was also used\footnote{Available at \url{http://on.ipleiria.pt/plenoisla}.}. The SKINL2 dataset, was acquired with a Raytrix R42 camera in a clinical setting in the context of dermatological research on skin lesions using LFs. The images used in this work were also submitted to the JPEG-Pleno standardisation initiative~\cite{JPEG_2020_JPEGPleno}. Figure~\ref{fig:dataset-examples} shows a SAI of one image of each of these datasets, and Table~\ref{tab:dataset-characteristics} lists their characteristics.
	
	The JPEG-Pleno toolchain uses the LF Toolbox~\cite{Dansereau_2013_CVaPR_DecodingCalibrationAndRectificationForLenseletBasedPlenopticCameras} to the decode the raw lenslet files from the Lytro dataset, \textit{i.e.} the EPFL dataset, and convert them to 4D LFs. The \textit{Colour Correction} step of the pre-processing chain expands the colour range which is detrimental to the coding efficiency, therefore, as the compression process is lossless, this mentioned step is omitted before encoding and if necessary performed at the decoder side.
	
	\begin{table}
		\centering
		\renewcommand{\arraystretch}{1.1}
		\caption{Characteristics of the datasets used in this work.}
		\label{tab:dataset-characteristics}
		\begin{tabular}{@{}lm{2.9cm}ll@{}}
			\toprule
			Dataset & Type                                            & Resolution    & Bit depth \\ \midrule
			EPFL    & Lenslet: acquired with a Lytro Illum B01 Camera & $ 13 \times 13 \times 625 \times 434 $ & 10        \\
			HCI & HDCA: computer generated & $ 9 \times 9 \times 512 \times 512 $ & 10\footnotemark \\
			SKINL2  & Lenslet: acquired with a Raytrix R42 camera             & $ 9 \times 9 \times 1920 \times 1080 $ & 10        \\ \bottomrule
		\end{tabular}
	\end{table}
	\footnotetext{Originally the HCI LFs have a bit-depth 8 bits but are provided as 10 bits in the JPEG-Pleno CTC dataset.}
	
	\begin{table*}
		\centering
		\caption{Encoding configurations of the state-of-the-art encoders used in this work.}
		\label{tab:encoders-cfgs}
		\begin{tabular}{@{}lm{3.5cm}lm{9.5cm}@{}}
			\toprule
			Encoder &
			Software &
			Input type &
			Configuration \\ \midrule
			HEVC &
			HM reference software v16.12, including the Format Range Extension (RExt)~\cite{Flynn_2016_HEVCRExt} &
			PVS using spiral scan &
			\begin{itemize}
				\item Main RExt profile (Profile=main-RExt)
				\item Lossless cost mode
				\item QP 0 (zero)
				\item Transform quantization bypass (TransquantBypassEnable=1)
				\item 32 intra period (IntraPeriod=32)
				\item 16 GOP size (GOPSize=16)
				\item CRA intra random access point (open GOP) (DecodingRefreshType=1)
				\item TZ Search motion estimation (FastSearch=1)
			\end{itemize} \\
			VVC &
			VTM reference software v10.0 &
			PVS using spiral scan &
			\begin{itemize}
				\item Using configuration file: Random Access GOP32
				\item Using lossless configuration files
				\item QP 0 (zero)
			\end{itemize} \\
			JPEG-LS &
			JPEG-LS reference software & Array of SAIs &
			\begin{itemize}
				\item Standard configuration (-ls 1)
				\item YCbCr colour transformation bypass active (-c)
			\end{itemize} \\
			CALIC &
			Martin Briano’s implementation~\cite{Briano_CALIC} & Array of SAIs &
			\begin{itemize}
				\item Lossless mode (-d=0)
				\item 2D CALIC (-m 1)
			\end{itemize} \\
			JPEG XL &
			JPEG XL reference software v0.7.0 & Array of SAIs &
			\begin{itemize}
				\item Lossless mode (-d 0)
				\item Maximum effort (-e 9)
				\item Parallel computations disabled (--num\_threads=0)
			\end{itemize} \\
			
			MRP &
			Author's software from~\cite{Santos_2016_PCSP_CompressionofMedicalImagesUsingMRPwithBiDirectionalPredictionandHistogramPacking} &
			PVS using spiral scan &
			\begin{itemize}
				\item Maximum number of classes dependent on image size
				\item Bidirectional prediction with GOPsize = 8 (-B -G 7)
				\item Histogram packing and variable block size prediction enabled
				\item Reference pixels: 56 pixels on the intra frame, 20 pixels on the current frame and 5 pixels on the reference frame for unidirectional prediction frames and 12 pixels on the current frame and 13 pixels on the reference frames for bidirectional prediction frames
			\end{itemize} \\ \bottomrule
		\end{tabular}
	\end{table*}
	
	The performances of the proposed encoding methods were compared with those of other state-of-the-art encoders like MRP, JPEG-LS~\cite{Weinberger_2000_LOCO-I, JPEG_1997_JPEGLS}, CALIC~\cite{Wu_1997_IToC_ContextBasedAdaptiveLosslessImageCoding, Briano_CALIC}, JPEG XL~\cite{Alakuijala_2019_JPEGXL}, HEVC~\cite{Sullivan_2012_HEVC, MPEG_2020_HEVC}, and VVC~\cite{Bross_2021_PI_VVC, MPEG_2020_VVC}. To ensure the uniformity of the experiments, the RCT is also applied to the input of the state-of-the-art encoders. The state-of-the-art encoders implementations and parametrisation are shown in Table~\ref{tab:encoders-cfgs}.
	
	\subsection{Comparison of proposals and state-of-the-art}
	
	The experimental results in Table~\ref{tab:coding-bpp} show the coding efficiency, measured in bits-per-pixel (bpp), \textit{i.e.}
	\begin{equation}
		bpp = \frac{B_{C_{1}} + B_{C_{2}} + B_{C_{3}}}{T \times S \times V \times U}    
	\end{equation}
	where  $B_{C_{i}}$ is the number of bits of the compressed colour component $ C_i$ (\textit{e.g.} YUV) of the image, $ T \times S $ represents the number of SAIs, and $ V \times U $ the number of pixels of a SAI. The three modes described in Section~\ref{sec:4d-mrp}, 4D-MRP, DT-4D-MRP, and M-MRP, have been designed to exploit the unique characteristics of light-field images. Their coding efficiencies were evaluated by encoding all the images listed before. The results are analysed and discussed next.
	
	\begin{table*}
		\centering
		\caption{Compression results (in bpp) for EPFL and HCI datasets.}
		\label{tab:coding-bpp}
		\begin{tabular}{@{}llccccccccc@{}}
			\toprule
			Type                  & Light Fields              & JPEG-LS & CALIC & JPEG XL & HM    & VVC   & MRP \cite{Santos_2016_PCSP_CompressionofMedicalImagesUsingMRPwithBiDirectionalPredictionandHistogramPacking}   & 4D-MRP & M-MRP  \cite{Santos_2019_DCC} & DT-4D-MRP      \\ \midrule
			\multirow{4}{*}{LL}   & Bikes                     & 16.42   & 15.94 & 16.07   & 13.69 & 13.72 & 12.73 & 11.43  & 11.47 & \textbf{11.34} \\
			& Danger\_de\_Mort          & 16.18   & 15.69 & 15.77   & 13.03 & 12.97 & 11.89 & 10.80  & 10.80 & \textbf{10.71} \\
			& Fountain\_and\_Vincent\_2 & 16.66   & 16.23 & 16.32   & 14.43 & 14.52 & 13.28 & 12.09  & 12.06 & \textbf{11.99} \\
			& Stone\_Pillars\_Outside   & 16.77   & 16.28 & 16.33   & 13.44 & 13.38 & 12.36 & 11.29  & 11.20 & \textbf{11.17} \\ \midrule
			\multirow{2}{*}{HDCA} & Greek                     & 7.25    & 7.17  & 6.54    & 5.55  & 5.36  & 5.40  & 4.82   & \textbf{4.78}  & \textbf{4.78}  \\
			& Sideboard                 & 12.19   & 12.17 & 11.04   & 8.17  & 8.09  & 8.43  & 7.66   & 7.62  & \textbf{7.58}  \\ \midrule
			\textbf{Average}      &                           & 14.24   & 13.91 & 13.68   & 11.39 & 11.34 & 10.68 & 9.68   & 9.66  & \textbf{9.59}  \\ \bottomrule
		\end{tabular}
	\end{table*}
	
	\subsubsection{Experiments using EPFL and HCI datasets}
	
	The results in Table~\ref{tab:coding-bpp} show that MRP surpasses standard codecs like JPEG, HEVC, or VVC. However, neither of these standard encoders, including MRP, is adapted to exploit the particular characteristics of LF representation data. Comparison of the proposed methods with the JPEG-LS, CALIC, JPEG XL, HEVC (HM), or VVC (VTM) coders, demonstrates the superiority of the former over the latter. Indeed the proposed 4D-MRP and DT-4D-MRP schemes surpass the compression of MRP by 1.00 to 1.09 bpp on average, respectively. This represents roughly 10\% bit-rate savings. Considering the most recent state-of-the-art image and video coding standard, VVC, the proposed methods present bit-rate savings of 15\%, on average, surpassing VVC (VTM) by 1.66 to 1.75 bpp, for 4D-MRP and DT-4D-MRP, respectively. To the best of the authors' knowledge DT-4D-MRP achieves the overall highest compression efficiency in the universe of state-of-the-art lossless LF encoders. Considering the intra only encoders, JPEG-LS, CALIC, and JPEG XL, it is clear that their performance is quite worse than the remaining encoders. This can be explained by the lack of tools to exploit inter SAI redundancy that the all the other codecs have in various ways.
	
	Regarding the individual evaluation of the three proposed methods, it can be seen that the DT-4D-MRP achieves the highest compression efficiency, both on average and for each image, by 0.6 to 0.9\%, when compared to the 4D-MRP and M-MRP, respectively. The dual quadtree prediction provides better adaptation of partitions to the image data, which in turn results in higher compression efficiency. The 4D-MRP encoder presents the lower coding efficiency of the three, which is explained by the highly granular partition used by the encoder. This is due to the simultaneous partitioning in all dimensions, which might not gather so much redundancies as possibly expected, also resulting in smaller blocks and more side information. M-MRP, which uses a 2D instead of 4D partitions, achieves an intermediate coding efficiency when compared with 4D-MRP and DT-4D-MRP. This is due to the fact that 2D partitions can use blocks that include pixels from multiples SAIs and also because larger blocks might be used, as splitting in the four-dimensions at once is not mandatory.
	
	\begin{table*}
		\centering
		\caption{Compression results (in bpp) for the SKINL2 dataset.}
		\label{tab:skinl2-coding-bpp}
		\begin{tabular}{@{}lccccccccc@{}}
			\toprule
			Light Fields     & JPEG-LS & CALIC & JPEG XL & HM   & VVC  & MRP \cite{Santos_2016_PCSP_CompressionofMedicalImagesUsingMRPwithBiDirectionalPredictionandHistogramPacking} & 4D-MRP & M-MRP \cite{Santos_2019_DCC}    & DT-4D-MRP \\ \midrule
			Img1             & 13.39   & 12.63 & 7.76    & 8.65 & 8.17 & 8.06 & 7.01   & \textbf{6.79} & 6.84      \\
			Img2             & 13.59   & 12.79 & 7.90    & 9.10 & 8.74 & 7.81 & 7.10   & 6.98 & \textbf{6.97}      \\
			Img3             & 13.61   & 12.82 & 7.80    & 9.18 & 8.79 & 8.00 & 7.23   & \textbf{7.07} & 7.09      \\ \midrule
			\textbf{Average} & 13.53   & 12.75 & 7.82    & 8.98 & 8.57 & 7.96 & 7.11   & \textbf{6.94} & 6.97      \\ \bottomrule
		\end{tabular}
	\end{table*}
	
	\subsubsection{Experiments using SKINL2 dataset}
	
	Further coding experiments using the SKINL2 dataset revealed that performance results are, generally, consistent with those observed with the non-medical images. The results shown in Table~\ref{tab:skinl2-coding-bpp} confirm the better performance of the the proposed methods, which achieve higher compression efficiency than the remaining state-of-the-art encoders, surpassing MRP by 0.85 to 1.02 bpp, for 4D-MRP and M-MRP, on average, which represent bit-rate savings larger than 11\%. In comparison with the VVC (VTM), the proposed methods present gains of 1.46 to 1.63 bpp, on average, for 4D-MRP and M-MRP, respectively, which represent up to 19\% of bit-rate savings. These results show that the proposed methods achieve a higher compression efficiency for the skin lesions dataset, when compared with the remaining encoders. This is a noteworthy conclusion, as in many application scenarios medical images are required to be losslessly compressed. Curiously, unlike what was shown in Table~\ref{tab:coding-bpp}, while JPEG-LS and CALIC still present the worst compression performance, JPEG XL is able to compete with the remaining encoders, despite not being able to exploit inter SAI redundancies. In particular, JPEG XL is only surpassed by the methods of the proposed unified native 4D framework, with savings ranging from 0.71 bpp to 0.88 bpp, or 9.1\% to 10.9\%. This might be explained by the group splitting of JPEG~XL, which divides larger images into sub images of $ 256 \times 256 $ pixels to be encoded separately, and the characteristics of the SKINL2 dataset.
	
	Regarding the comparison between the three proposed methods, it can be observed that DT-4D-MRP and M-MRP achieve similar compression efficiency, with a slight advantage for M-MRP. For this dataset (SKINL2) the bit-rate savings of M-MRP are 2\% and 0.43\% when compared with 4D-MRP and DT-4D-MRP, respectively. This shows that there is a relative loss of efficiency of 4D-MRP and DT-4D-MRP for the current dataset, when compared with the results of Table~\ref{tab:coding-bpp}. This might be explained by the larger resolution of the images in the SKINL2 dataset, that seems to favour the M-MRP algorithm.
	
	Overall the experimental evaluation demonstrates, in a consistent manner, that using both 4D prediction and 4D partition to encode LFs in MRP, leads to the higher coding efficiency of DT-4D-MRP.
	
	\subsection{Analysis of computational complexity}
	
	\begin{table*}
		\centering
		\caption{Encoding time ratio in comparison with MRP, for EPFL and HCI datasets}
		\label{tab:coding-times}
		\begin{tabular}{@{}llccccccccc@{}}
			\toprule
			Type                  & Light Fields              & JPEG-LS & CALIC  & JPEG XL & HM  & VVC & MRP \cite{Santos_2016_PCSP_CompressionofMedicalImagesUsingMRPwithBiDirectionalPredictionandHistogramPacking} & 4D-MRP & M-MRP \cite{Santos_2019_DCC} & DT-4D-MRP \\ \midrule
			\multirow{4}{*}{LL}   & Bikes                     & 1.5E-3  & 1.0E-2 & 0.2     & 0.2 & 3.6 & 1.0 & 3.5    & 6.4   & 91.9      \\
			& Danger\_de\_Mort          & 2.0E-3  & 1.4E-2 & 0.2     & 0.3 & 4.4 & 1.0 & 4.7    & 9.3   & 90.2      \\
			& Fountain\_and\_Vincent\_2 & 1.9E-3  & 1.1E-2 & 0.2     & 0.3 & 4.6 & 1.0 & 3.7    & 8.8   & 122.5     \\
			& Stone\_Pillars\_Outside   & 1.7E-3  & 1.1E-2 & 0.2     & 0.3 & 4.5 & 1.0 & 3.7    & 8.0   & 89.0      \\ \midrule
			\multirow{2}{*}{HDCA} & Greek                     & 8.9E-4  & 6.2E-3 & 0.1     & 0.1 & 3.0 & 1.0 & 4.9    & 6.6   & 35.1      \\
			& Sideboard                 & 1.0E-3  & 6.4E-3 & 0.1     & 0.2 & 2.6 & 1.0 & 4.3    & 4.1   & 44.4      \\ \midrule
			\textbf{Average}      &                           & 1.5E-3  & 9.8E-3 & 0.2     & 0.2 & 3.8 & 1.0 & 4.1    & 7.1   & 78.9      \\ \bottomrule
		\end{tabular}
	\end{table*}
	
	Besides coding efficiency, the computational complexity is an important benchmarking aspect, although not always primary. The complexity results pertaining to the different encoders, evaluated by encoding the EPFL and HCI dataset images, are presented in Table~\ref{tab:coding-times}. These results were obtained on a server running Ubuntu Server 20.04.1 LTS, equipped with an Intel Xeon(R) Silver 4114@2.20GHz CPU, and 192GB of DDR4@2666MHz RAM. The results show that the DT-4D-MRP has the worst performance in terms of complexity, when compared with the other encoders, taking on average 78 times longer than MRP to encode the same images. As stated in Section~\ref{sec:dt-4d-mrp}, this high computational complexity is due to the large number of partitioning options that need to be tested in the dual quadtree. The 4D-MRP and M-MRP proposals have computational complexities of the same magnitude, about 4 and 7 times that of MRP respectively. However they have a smaller compression performance than that of DT-4D-MRP. Despite VVC presenting about the same complexity as the 4D-MRP and M-MRP encoders, its compression performance is much lower than that of these encoders. Its performance is on pair with that of HEVC, despite being an older standard. This might be explained due to the developing effort of VVC having been so far focused on lossy compression performance with smaller improvements to the lossless modes.
	
	\vfill
	
	The three prediction modes described in Section~\ref{sec:4d-mrp} increase the compression efficiency of MRP encoding and to the best of the authors' knowledge, DT-4D-MRP presents the best lossless performance for LF images. While its computational complexity might be a concern, some adjustments can be made to deal with it. The authors have shown in~\cite{Santos_2019_EELFIWE_RateComplexity} that the complexity of M-MRP can be lowered, while still maintaining better compression efficiency than other state-of-that-art encoders. Additionally, none of the computer code of the proposed methods was optimised for computational efficiency, which leaves room for improvements. In the case of 4D-MRP and M-MRP, while they present lower performance than DT-4D-MRP, still have competitive coding ratios and can be regarded as a good compromise in terms of rate-complexity trade-off. Most of the computational complexity of the proposed framework lays at the encoder side, with the three methods having about the same computation times for the decoder. Given these considerations, the DT-4D-MRP might be a good candidate for offline compression. One application is related to large databases, such as in medical facilities, where the compression efficiency is more important than the computation times. In this case, any amount of extra compression can lead to a more efficient use of the available storage capacity. On the other hand, when computational efficiency is a concern, for instance for domestic users, one of the other prediction modes can be chosen.

	\section{Conclusion and Future Work}\label{sec:conclusion}
	
	\subsection{Conclusion}
	
	In this paper, three 4D prediction and partition modes for LF lossless encoding are studied. These allow the encoder to use up to four neighbouring SAIs as prediction references, which leads to better prediction performance and increases the coding efficiency of the MRP encoding, by exploiting the characteristics of LFs. The three types of prediction used to select the best predictors use 4D partitions with (i) an hexadeca tree (4D-MRP), (ii) a 4D partition using a quadtree for the spatial dimensions and other for the angular dimensions (DT-4D-MRP), and (iii) a conventional 2D partition (M-MRP) with multiple references, previously described in~\cite{Santos_2019_DCC}. The 4D-MRP and M-MRP prediction modes present a good compromise in terms of rate-complexity trade-off.
	
	Overall, in comparison with other state-of-the-art encoders, all the proposed methods present higher compression ratios, reducing the average bit rates from 10 to 32\%. The highest performance is achieved by DT-4D-MRP which to the best of the authors' knowledge is the best lossless coding performance for LFs represented in the 4D spatial-angular domain achieved so far.
	
	\subsection{Future Work}
	
	A few research directions are left open by this paper and are intended to be addressed in the future. First and foremost, the authors aim to address the computational complexity of DT-4D-MRP, both by proper selecting the input parameters, as was done in~\cite{Santos_2019_EELFIWE_RateComplexity} and also through a comprehensive analysis of the exact causes of this encoder computational complexity, aiming to propose directed solutions for this issue.
		
	A previous work has shown that the bulk of the MRP bitstream is used to represent the prediction residuals~\cite{Santos_2018_JoVCaIR_LosslessCodingofLightFieldImagesBasedonMinimumRatePredictors}. Therefore, it should be worth it to investigate methods to improve the arithmetic encoding of these residuals. One idea that has not yet been explored is to forgo the use of a discretised set of probability models in favour of more modern methods, such as the adaptive learning of the probability models from previously decoded symbols.
	
	\ifCLASSOPTIONcaptionsoff
	\newpage
	\fi
	
	
	
	\bibliographystyle{bibliography/IEEEtran}
	\bibliography{bibliography/IEEEabrv.bib,bibliography/bibliography.bib}

\end{document}